\documentclass[preprint,11pt,a4wide]{elsarticle}

\usepackage[latin1]{inputenc} 
\usepackage{leqno}
\usepackage{amssymb}
\usepackage{verbatim}
\usepackage{xspace}
\usepackage{marvosym}
\usepackage{subfig} 
\usepackage{graphicx}
\usepackage{hyperref} 
\usepackage{txfonts}
\usepackage{multirow}
\usepackage{rotating} 
\usepackage{booktabs}
\usepackage{longtable}

\usepackage{dcolumn} 
\newcolumntype{E}{D{.}{.}{2,11}}
\newcolumntype{e}{D{.}{.}{2,8}}
\newcolumntype{F}{D{.}{.}{2,5}}
\newcolumntype{f}{D{.}{.}{2,4}}
\newcolumntype{T}{D{.}{.}{2,3}}

\graphicspath{{./EmFigures/}}

\def\eqb{\begin{equation}}
\def\eqe{\end{equation}}
\def\itb{\begin{itemize}}
\def\ite{\end{itemize}}
\def\enb{\begin{enumerate}}
\def\ene{\end{enumerate}}

\def\lbra{\left (}
\def\rbra{\right )}

\newcommand{\unit}[1]{\,\mathrm{#1}}
\newcommand{\cher}{\xspace{Cherenkov}\xspace}
\newcommand{\cherl}{\xspace{Cherenkov-light}\xspace}
\newcommand{\chera}{\xspace{Cherenkov-angle}\xspace}
\newcommand{\cherthr}{\xspace{Cherenkov-threshold}\xspace}

\newcommand{\cherphos}{\xspace{Cherenkov-photons}\xspace}

\newcommand{\tamm}{\xspace{Frank-Tamm}\xspace}
\newcommand{\tammf}{\xspace{Frank-Tamm factor}\xspace}

\newcommand{\geantdrei}{\xspace{Geant3.16}\xspace}
\newcommand{\geantfour}{\xspace{Geant4}\xspace}

\newcommand{\ud}{\mathrm{d}}

\journal{Astroparticle Physics}

\begin{document}

\begin{frontmatter}
\title{{\bf Calculation of the \cherl yield from electromagnetic cascades in ice with Geant4\tnoteref{t1}}}
\tnotetext[t1]{published in Astroparticle Physics 44 (2013) 102 113}

\author[rwth]{Leif~R\"adel}
\ead{Leif.Raedel@physik.rwth-aachen.de}

\author[rwth]{Christopher~Wiebusch}
\ead{Christopher.Wiebusch@physik.rwth-aachen.de}

\address[rwth]{III.Physikalisches Institut, Physikzentrum, RWTH Aachen University, Otto Blumenthalstrasse, 52074 Aachen, Germany}


\date{Astroparticle Physics 44 (2013) 102 113}


\begin{abstract}

In this work we investigate and parameterize the amount and angular
distribution of Cherenkov photons which are generated by 
electromagnetic cascades in water or ice. 
We simulate electromagnetic cascades with Geant4 for primary electrons, positrons and photons with energies ranging from 1 GeV to 
10 TeV. We parameterize the total Cherenkov light yield as a function of energy,
the longitudinal evolution of the Cherenkov emission along the cascade-axis and the angular distribution of photons. Furthermore, we investigate the fluctuations of the total light yield, the fluctuations in 
azimuth and changes of the emission with increasing age of the cascade.

\end{abstract}

\end{frontmatter}

\section{Introduction}

High-energy neutrino telescopes such as IceCube, 
Baikal or Antares \cite{ICECUBE,BAIKAL,ANTARES} 
detect \cherl from charged  
particles  in natural media like water or ice.
\cherl is produced when these particles propagate through the medium
with a  speed faster than the phase velocity of light $ v> c_{med} = c/n $.
In ice and water  the  refraction index $n$ is typically 
$n \approx 1.33$ \cite{KUZMICHEV,PRICEGROUP}. Hence, the \cherthr is given by  $\beta = \frac{1}{n} $
which corresponds to a minimum kinetic energy of 
\eqb \label{eq:thrthr}
E_{c} = m \cdot \lbra \frac{1}{\sqrt{1-{1\over n^2}}} -1\rbra ~.\eqe
 For electrons this is 
$E_{c} \approx 0.26$\,MeV  in water and ice.
The number of emitted photons per unit track and wavelength interval
 is given by the \tamm formula \cite{PDG,frank1937cerenkov} 
\eqb \label{eq:tamm}
{d^2 N \over dx d\lambda } = {2 \pi \alpha z^2 \over \lambda^2 }\cdot 
sin^2 (\theta_c ) ~.
\eqe
Here $\theta_c $ is the \chera. This is the opening angle of a cone
into  which the photons are emitted
\eqb \label{eq:cerangle}
\cos{(\theta_c)} = \frac{1}{n\beta} ~.
\eqe
A 
relativistic track ($\beta = 1$) in water or ice ($n \approx 1.33$)  produces
about  $N_0 \approx 250 \, \mathrm{cm}^{-1} $ optical photons
in a wavelength interval between $300$\,nm and
 $500$\,nm, which is a typical sensitive region of photo-detectors, e.g. \cite{ICPMT}, used in 
the aforementioned neutrino telescopes.
The \chera for a relativistic track ($\beta =1 $) in ice is  
$ \theta_{c,0} = \arccos (1/n)  \approx 41^\circ $.

A large fraction of detected \cherphos in high-energy neutrino telescopes 
originates from electromagnetic cascades. These
are initiated by a high-energy electromagnetic particle which produces
a shower of secondary  particles by subsequent bremsstrahlung and  pair production
processes \cite{PDG,heitler1954quantum}.
The primary particle can  originate from radiative energy losses of 
a high-energy muon
(bremsstrahlung and pair production), or from the decay of $\pi^0\to 2 \gamma $
 in hadronic cascades, or be a high-energy electron from a charged-current 
interaction of an electron neutrino.

\begin{figure}[htp]
    \includegraphics*[width=.95\textwidth]{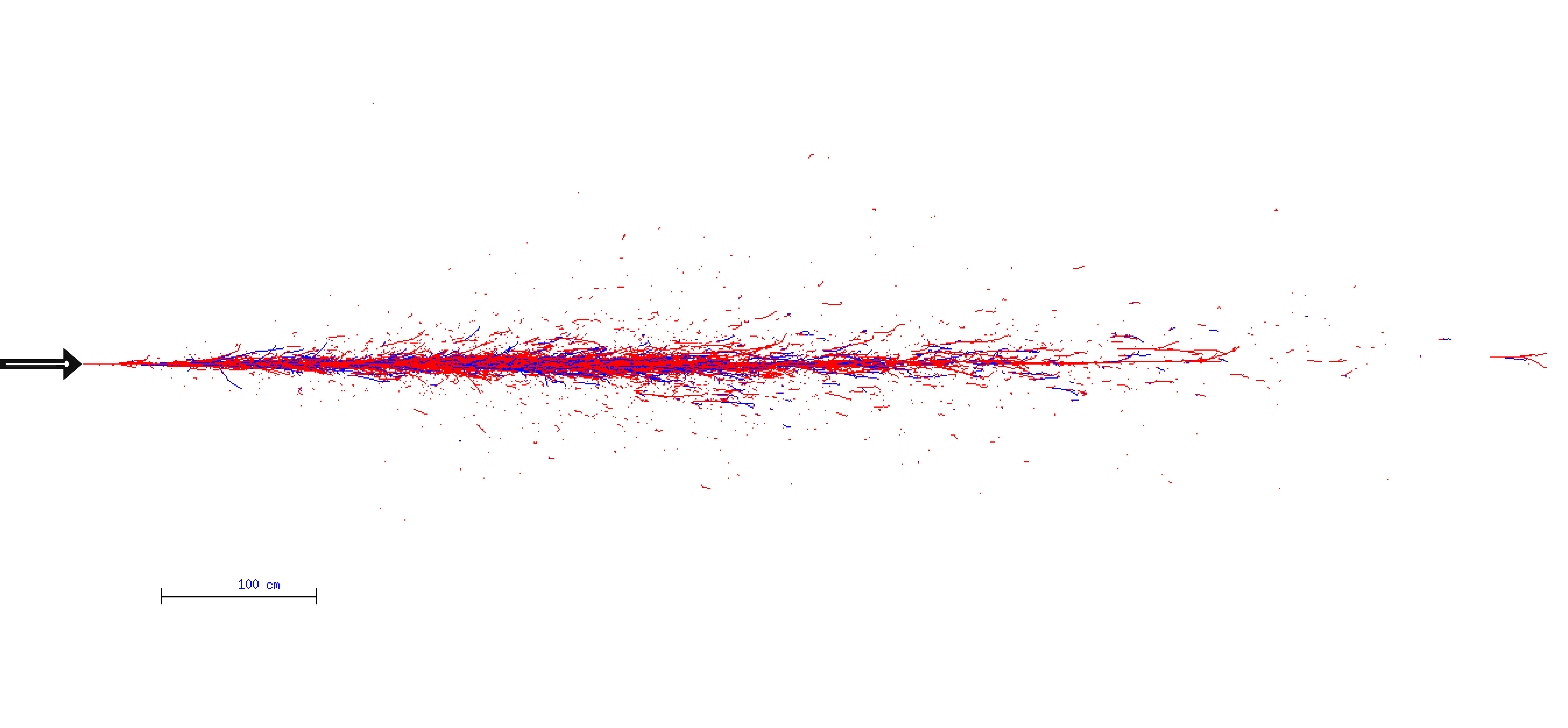}
    \caption{A simulated electromagnetic cascade. A primary electron of  
$100\unit{GeV}$ has been injected at the left pointing towards the right. Shown are all generated charged secondary particles (red for negative and blue for positive charge) as the result of a \geantfour simulation.
Neutral particles, like photons are not shown. 
     \label{fig:cascade:exam}}
\end{figure}

An example of a simulated cascade is shown in figure \ref{fig:cascade:exam}.
Each particle in the cascade produces \cherl according to Eqs.(\ref{eq:tamm})
 and (\ref{eq:cerangle}), if its energy is above the \cherthr 
Eq.(\ref{eq:thrthr}). Due to multiple interactions and scattering 
the directions of the particles in the  cascade 
 differ from that of the original primary particle
and a broad angular distribution of emitted \cherphos is expected.

The characteristic length scale for the development of an electromagnetic 
cascade
is given by the radiation length $X_0$ \cite{PDG}.
It is  about
$X_{0,ice} \approx 39.75\unit{cm}$  and $X_{0,water} \approx 36.08\unit{cm}$ as determined by  \geantfour for the configuration  listed in \ref{app:geant:conf}.
The length  along the shower axis $z$ is usually expressed by the  dimensionless shower depth 
\eqb
t \equiv z/X_0  ~.
\eqe
The  length of the shower increases typically 
logarithmically with the ratio of primary energy $E_0 $ and critical energy $E_{crit} $. The critical energy is the energy above which radiative process dominate the energy loss of electrons. The values for ice and water obtained from
\cite{PDGatomicproperties}  are
 $E_{crit,ice}^{\mathrm{e-}} = 78.60\unit{MeV}$, $E_{crit,ice}^{\mathrm{e+}} = 76.51\unit{MeV}$
and $E_{crit,water}^{\mathrm{e-}} = 78.33\unit{MeV}$, $E_{water,ice}^{\mathrm{e+}} = 76.24\unit{MeV}$. 

The physical length of a shower 
is typically less than $10$\,m. This is short, compared to the scale of neutrino telescopes and the full \cherl is created locally and expands with 
time as an almost spherical shell  with a 
characteristic angular  distribution of the intensity.

Due to the large number of particles the full tracking of each particle
in Monte-Carlo simulations of cascades  in neutrino telescopes is very time 
consuming. However, the development of electromagnetic
cascades is very regular because fluctuations are statistically suppressed by the large number of interactions
and large number of involved particles. Hence, it can be  well 
approximated by the average development.
Therefore, for the simulation of data in neutrino telescopes, 
 the average \cherl output can be parameterized, e.g. as done
 in \cite{CHWPHD,KOWALSKI,KAHAE,MIRANI}.

This work follows up the work in \cite{CHWPHD} which was based
on \geantdrei \cite{GEANT316} with a more precise calculation 
of the total Cherenkov-light yield and its angular distribution based on \geantfour
\cite{GEANT4}.
For different primary energies and primary particles,
 we investigate  the velocity distribution and the directional 
distribution of particles and the longitudinal development of the cascade.
We present a parameterization of the \cherl yield and investigate
its fluctuations as well as variations of the azimuthal 
symmetry of the cascade.
We also  present a parameterization of the angular distribution of \cherphos
and investigate variations of this distribution during the development of the cascade. The results are compared to \cite{CHWPHD,KOWALSKI,KAHAE,MIRANI}.

We note, that similar calculations have been also been performed for the calcualtion of coherent radio emission from electro-magnetic cascades in ice \cite{razzaque2002coherent}. Although those calculations do not consider the photon yield of Cherenkov light, and concentrate on the radio emission their results e.g. for the total track-length are similar. 

\section{Simulation method \label{sec:mcframe}}

The calculation of this work  follows largely the strategy described  
in \cite{NAKED}.
We use the  \geantfour (GEometry ANd Tracking) toolkit to track 
the particles in the cascade through the medium ice or water \cite{GEANT4}. 
 The used media properties are given in  \ref{app:geant:conf}.
Unless noted otherwise, we used an index of refraction of  $n=1.33 $  and a density of   $\rho_{ice}= 0.91\unit{g/cm^3} $.
Note, that these values   slightly deviate 
from the values in  \cite{PDGatomicproperties}:  $\rho_{ice}= 0.918\unit{g/cm^3} $ and  $n_{ice} = 1.31 $ and the value 
 $\rho_{ice}= 0.9216\unit{g/cm^3}  $ at the center of IceCube \cite{DIMA}. This introduces a small systematic uncertainty of about $1\% $, which can be corrected for by rescaling
our results to the correct density.

\begin{figure}[htp]
    \centering
    \includegraphics*[width=.45\textwidth]{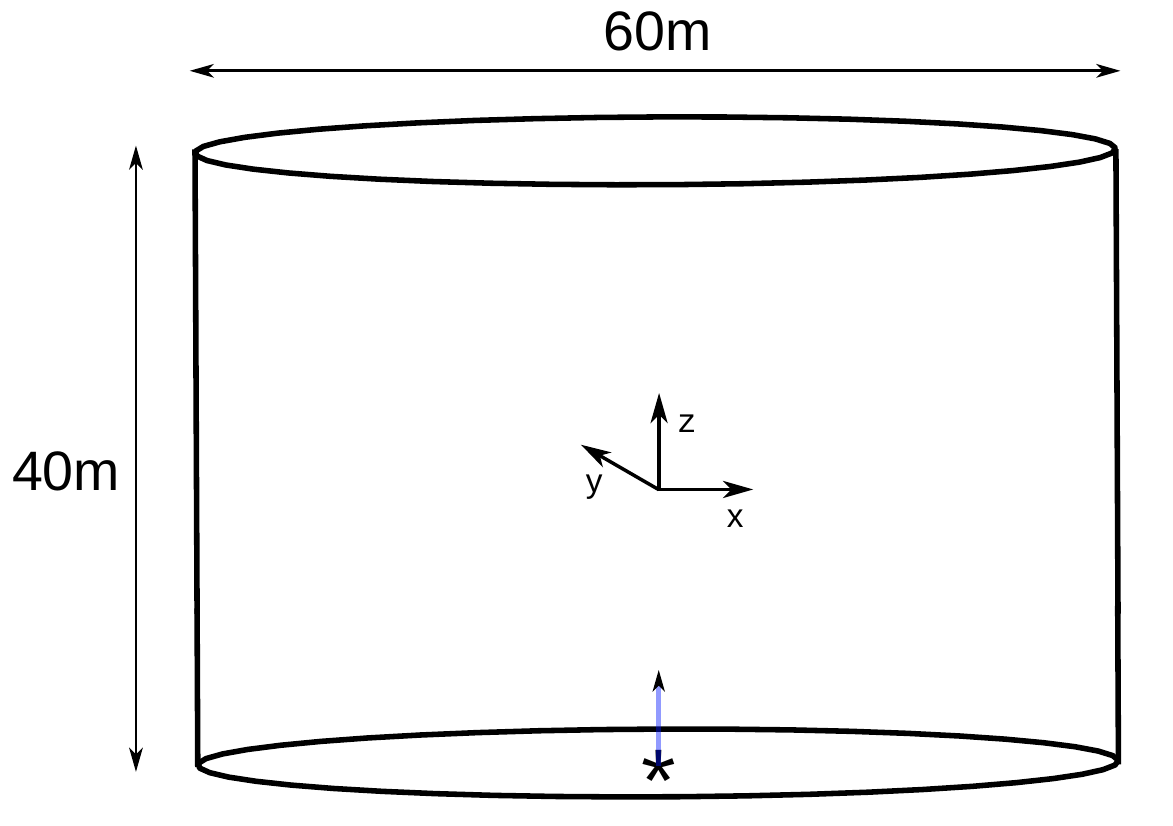}
    \hfill
    \includegraphics*[height=.45\textwidth,width=.45\textwidth]{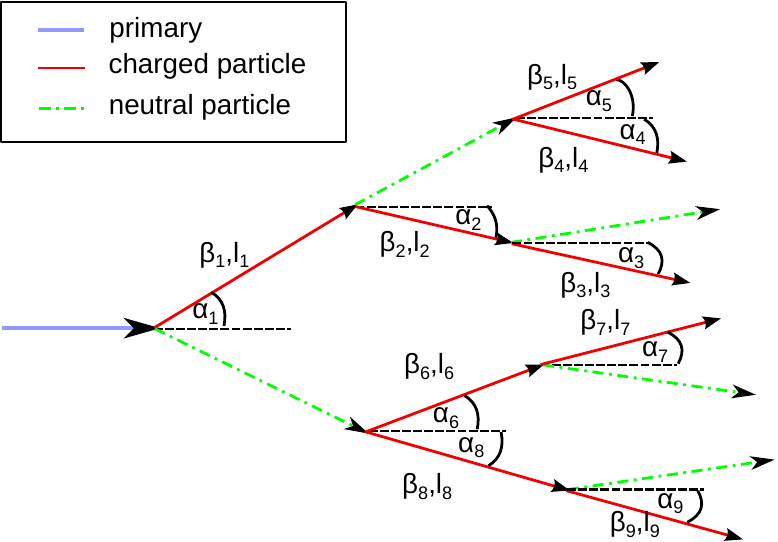}
    \caption{Geometry of the simulation and method of the calculation
     \label{fig:method}}
\end{figure}

The  simulation principle of this work is illustrated in 
figure \ref{fig:method}. 
The medium is contained in a cylindrical volume of $30$\,m radius and 
$40\unit{m} $ height.  The dimensions
are chosen such that all secondary particles are well confined  within the 
geometry and  fully tracked.
The  primary particle $e^\pm $ or $\gamma $  is injected at 
 the bottom center into this volume
with its initial momentum pointing into positive z-direction.
The particles are propagated through the medium and secondary 
particles are created, which again can produce further particles. 
Each step between two interactions
corresponds to a track segment for which the energy and direction
are assumed constant.
For each  track segment $i$ we store the length $l_i$, the Lorentz factor $\beta_i $, the $z$-position $z_i$ and the direction $\alpha_i$ with respect to the
z-axis. The azimuth angle $\phi $ discussed in section \ref{sec:azi} corresponds to the rotation angle in the x-y plane.
Summing over all track segments allows to calculate the 
Cherenkov photon yield and the corresponding angular distribution.

For the here described simulation it is important to simulate all 
particles with energies above the 
\cherthr. Details on the simulated physics processes are given 
in \ref{app:geant:conf}.
 In \geantfour some electromagnetic processes require production thresholds to avoid infrared divergences \cite{GEANT4UserGuideApplication}. 
These production thresholds are specified as a cut-in-range threshold, 
using the  \texttt{SetCuts()} method of \texttt{G4VUserPhysicsList}.  
Here,  particles are  tracked if their mean expected range is 
larger than this cut-in-range threshold.
For each material and particle type, this  cut-in-range is 
transformed into a corresponding energy threshold.
Here, a cut-in-range of $100\,\mu\mathrm{m}$ is chosen. 
This corresponds to a kinetic energy threshold
of $E_{\mathrm{cut},e^{\pm}} \approx 80\unit{keV}$ for electrons, 
which is well below the \cherthr 
$E_{\mathrm{c},e^{\pm}}\approx 264\unit{keV}$. 
Once produced, all secondary particles are tracked until 
they stop. In order to increase the computing performance, a single 
scattering process of a particle does not correspond to an individual 
track segment  but  multiple scattering processes are 
simulated as  one step.

We perform simulations up to a maximum energy of the primary particle
of  $10$\,TeV to which electromagnetic   cross sections in \geantfour 
are valid. 
Our results could be extrapolated beyond this limit,
however, at high energies  an additional effect, the LPM effect,
is expected to set in. This effect describes significantly reduced
 electromagnetic cross sections and the longitudinal development 
of such cascades would become strongly elongated such that our 
parameterization approach is not valid.

For $\beta=1 $, the number of emitted \cherphos is proportional to the length
of the track and can be calculated using
Eq.(\ref{eq:tamm}). For  $\beta<1$ the photon yield is smaller and proportional
 to the factor
\eqb
\sin^2 (\theta_c ) = 1- \cos^2 (\theta_c ) = 1-\frac{1}{\beta^2 \cdot n^2}  ~.
\eqe
In order to properly account for this smaller yield, the length of each track segment $l$ is  scaled with the \tammf
\eqb \label{eq:tamm:factor}
\hat{l} = \frac{\sin^2 (\theta_c )}{sin^2(\theta_{c,0}) } \cdot l
\quad \mbox{with} \quad
\sin^2 ( \theta_{c,0} ) = 1-\frac{1}{n^2} ~.
\eqe
The value $\hat{l} $ thus corresponds to the equivalent length of a 
relativistic track with the same photon yield as the track length $l$.
The use of the equivalent length $\hat{l} $ instead of an explicit 
calculation of photons has the advantage that the here presented
results can be rescaled 
 to slightly different indices of refraction, and are independent of the assumed wavelength interval of the considered photo-detector.

The angular distribution of the \cherphos is
calculated with the method introduced in \cite{NAKED}.
In this method the distribution of track length as a function
of the directional angle $\alpha $  
with respect to the z-axis (zenith) and the velocity $\beta $
can be transformed to a zenith distribution of emitted \cherphos.
The prerequisite for the applicability of that method is a high statistics of tracks 
which are distributed uniformly in azimuth.

\section{Results}

\subsection{Velocity distribution of shower particles}

\begin{figure}[!htp]
    \centering
    \includegraphics*[width=.58\textwidth]{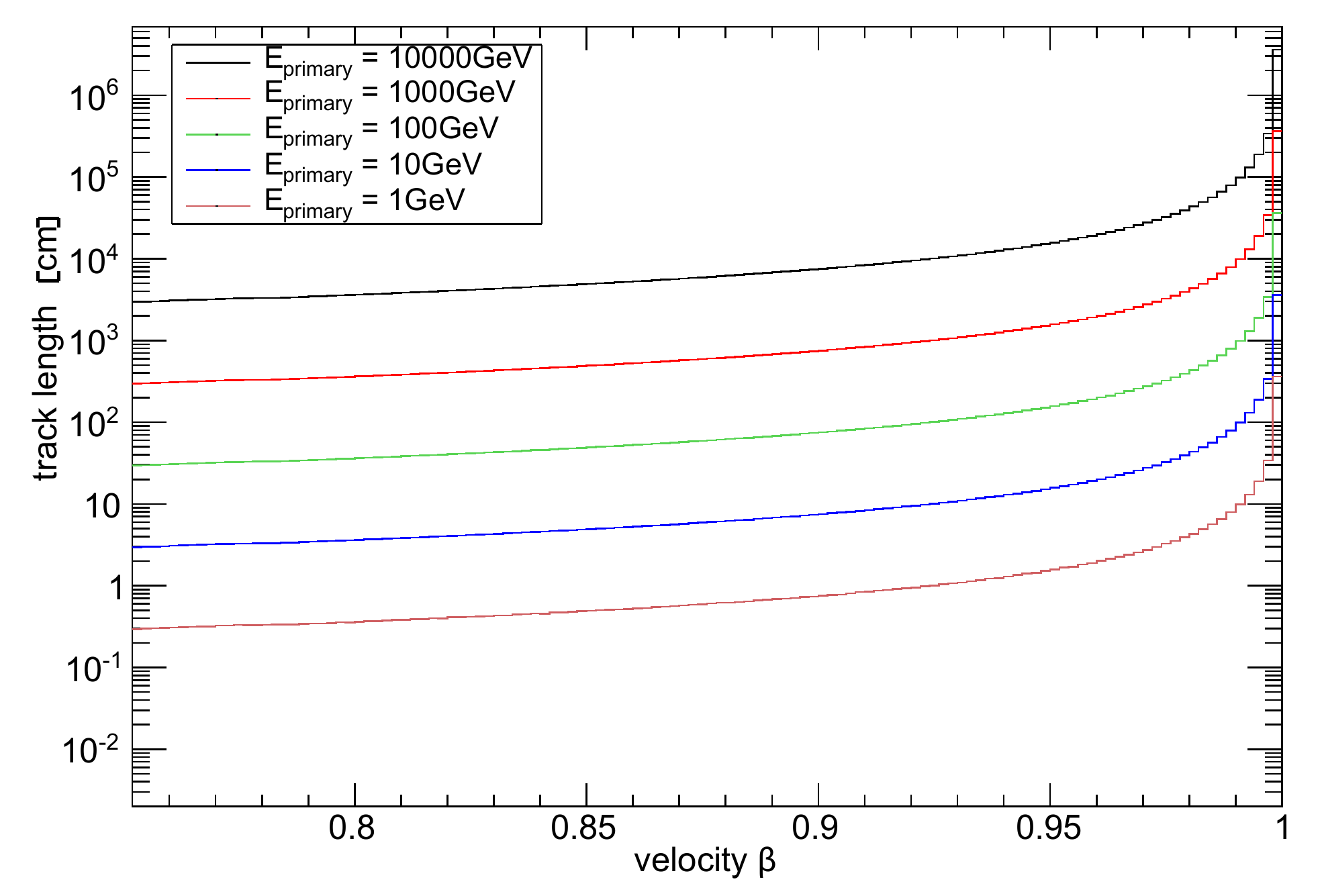}
    \includegraphics*[width=.58\textwidth]{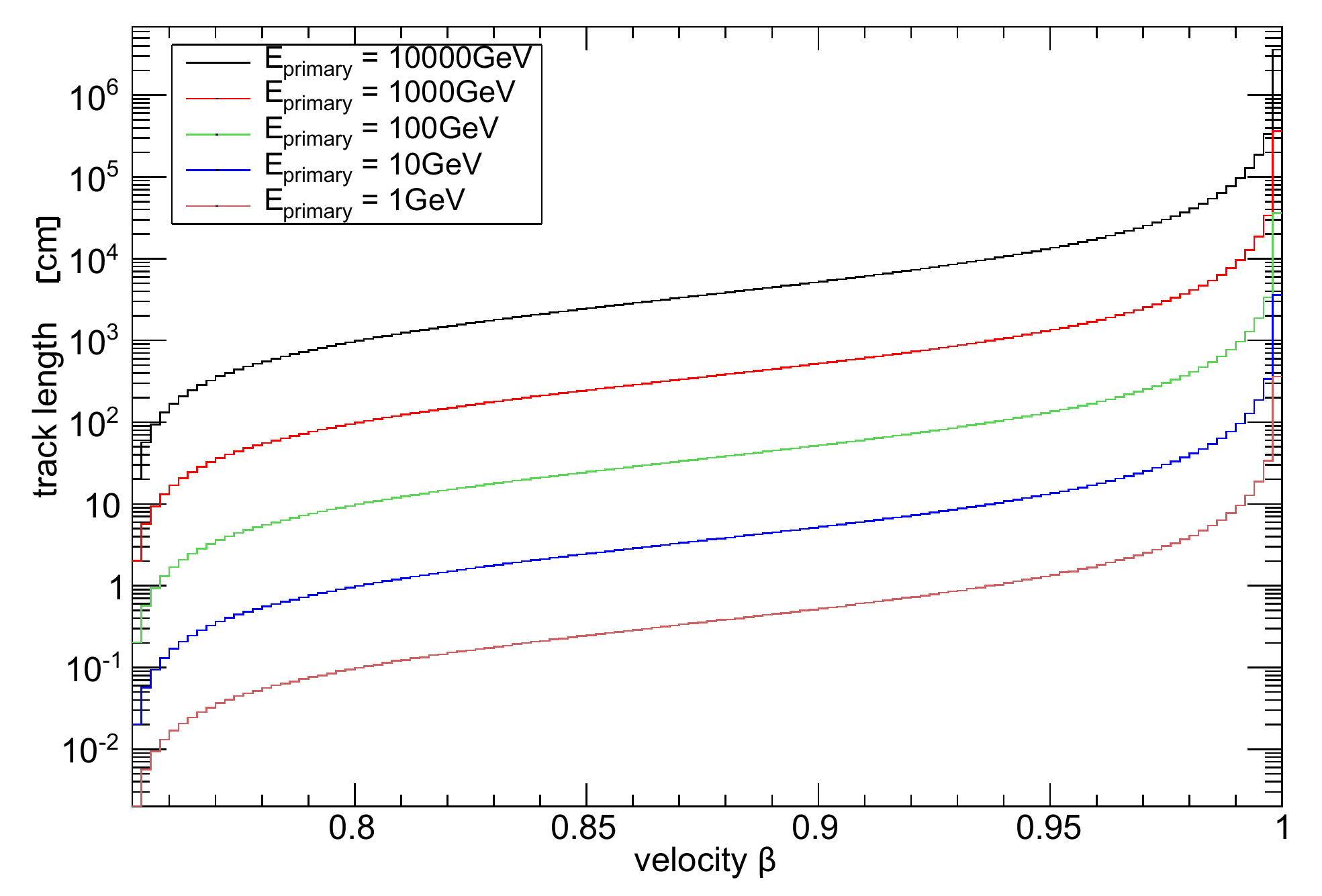}
    \includegraphics*[width=.58\textwidth]{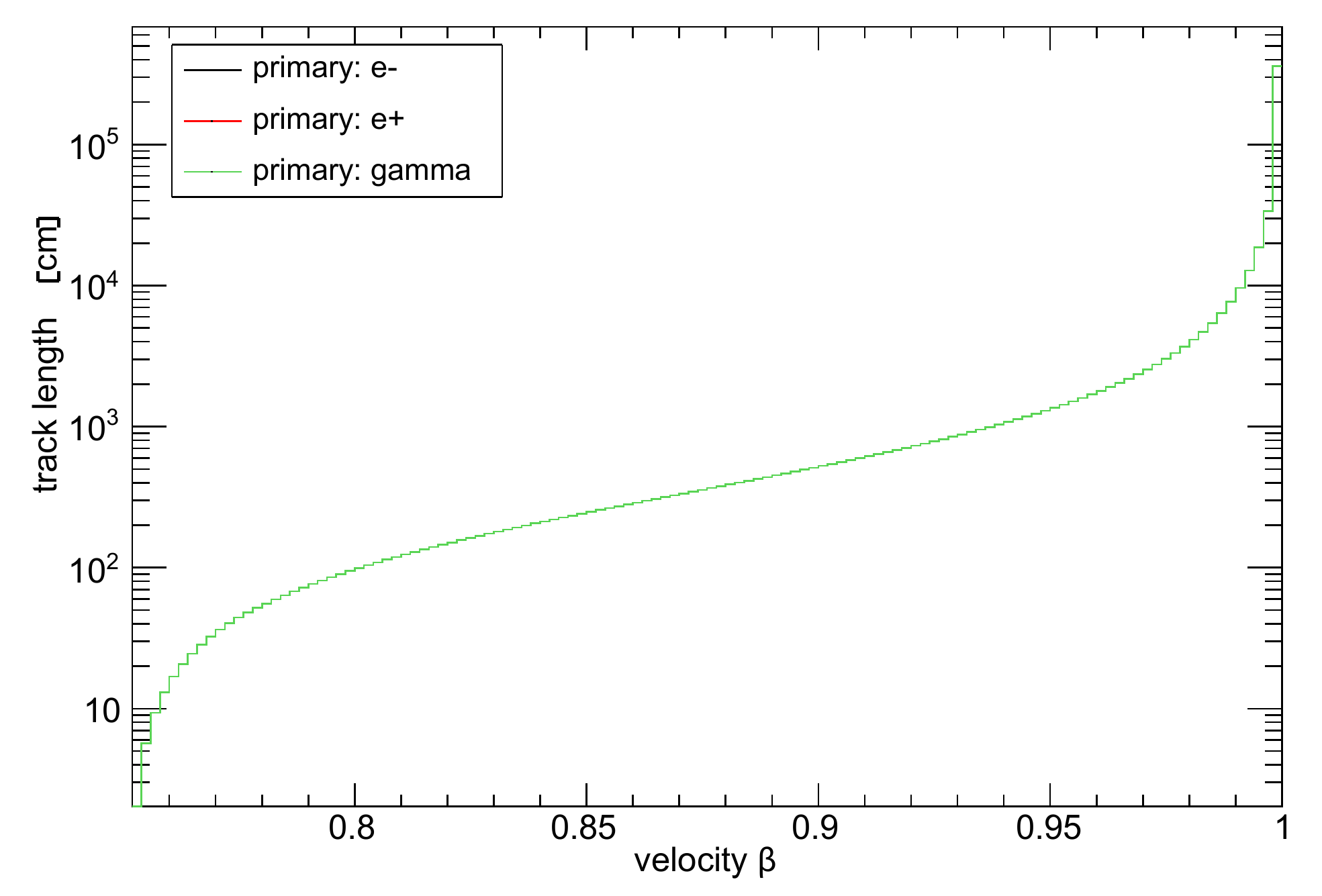}
    \caption{Velocity distribution of  track length. 
Shown is the differential distribution of summed track length per shower versus 
the Lorentz factor $\beta$ for bins of 0.002 in $\beta$. The top figure 
show the distributions of physical length $l$ for the shower from primary  
positrons of  different   energies. The middle figure shows the same distribution for the track length $\hat l$ which has been weighted with the \tammf, Eq.(\ref{eq:tamm:factor}).
The bottom figure shows the distributions of $\hat l$ for  
different primary particles: $e^+ $, $e^- $, $\gamma $ for the  
        primary energy $E_{0}=1\unit{TeV}$.
     \label{fig:beta}}
\end{figure}

Figure \ref{fig:beta} shows the distribution of track length 
in the cascade versus the Lorentz factor $\beta $ for different 
primary energies $E_0$. The shape is remarkably constant for different $E_0$
 while the total normalization is proportional to $E_0$.
The distribution also does not depend on the type of the primary particle.
The difference between the physical track-length $l$ and the effective track length $\hat l$ becomes particularly obvious close to the \cherthr $\beta \approx 0.752 $. Unless noted otherwise, we will use in the following 
the effective track length $\hat l $ instead of the physical track length $l$.

\subsection{Total light yield \label{sec:res:total}}

\begin{figure}[htp]
    \centering
    \includegraphics*[width=.58\textwidth]{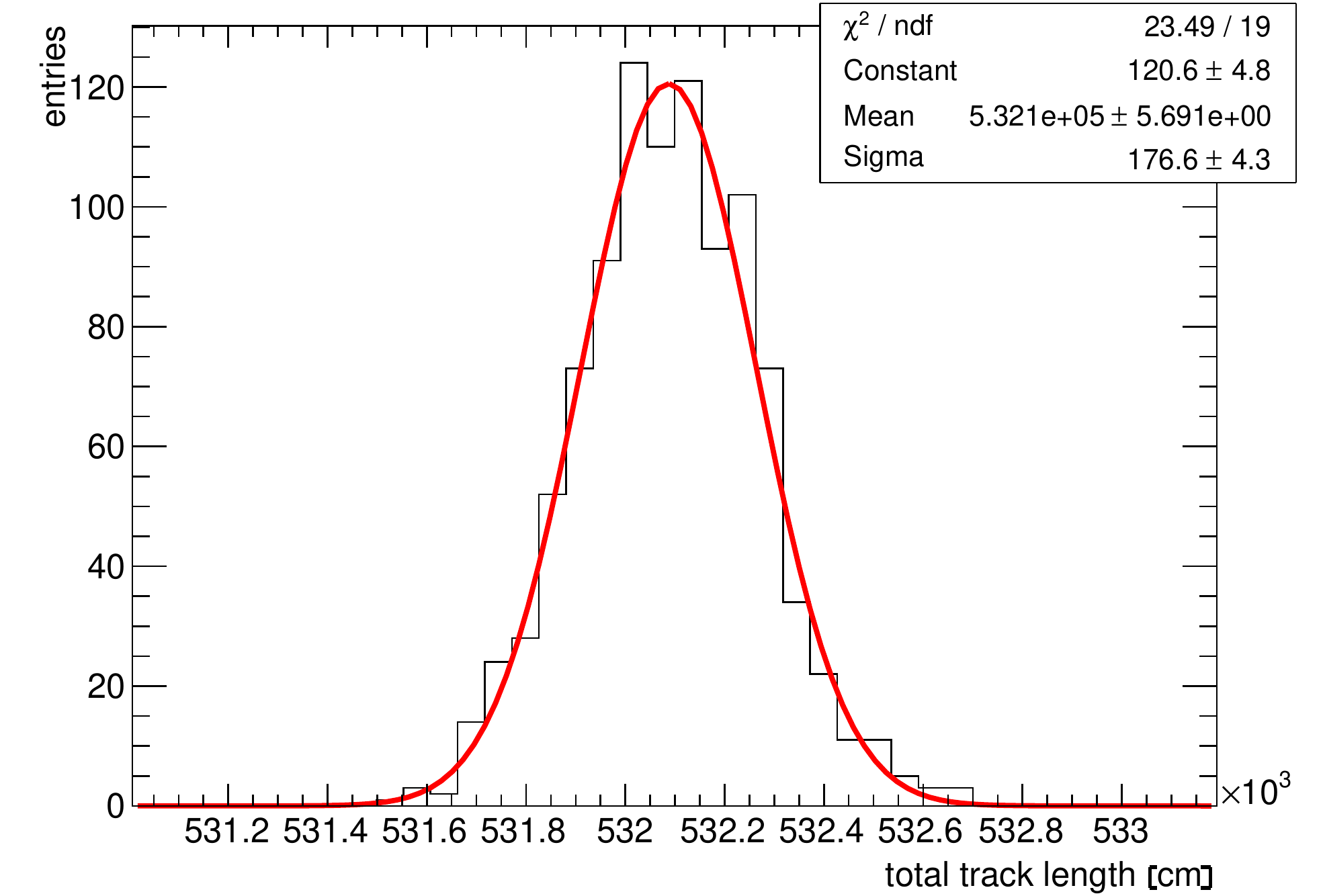}
    \includegraphics*[width=.58\textwidth]{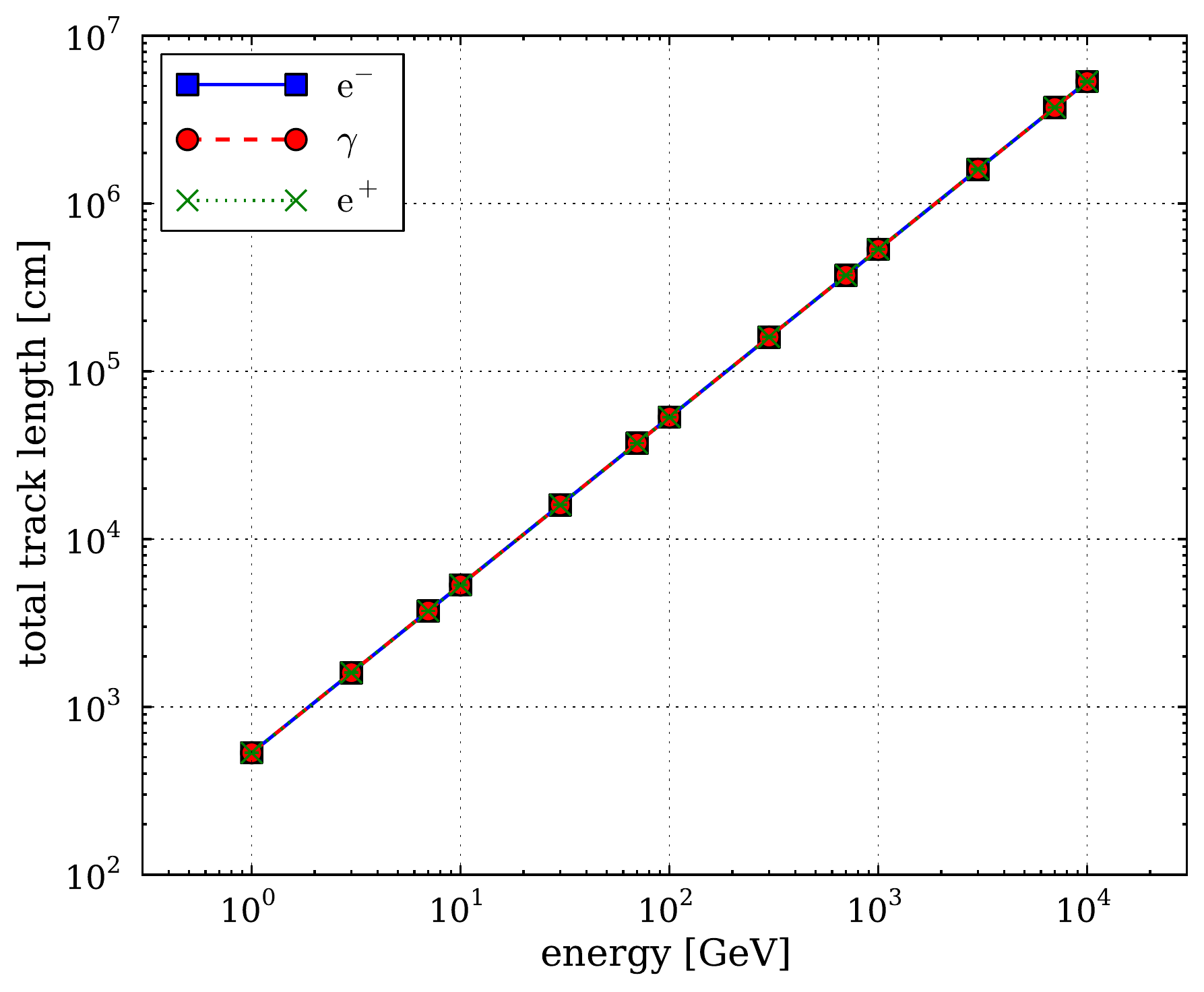}
    \includegraphics*[width=.58\textwidth]{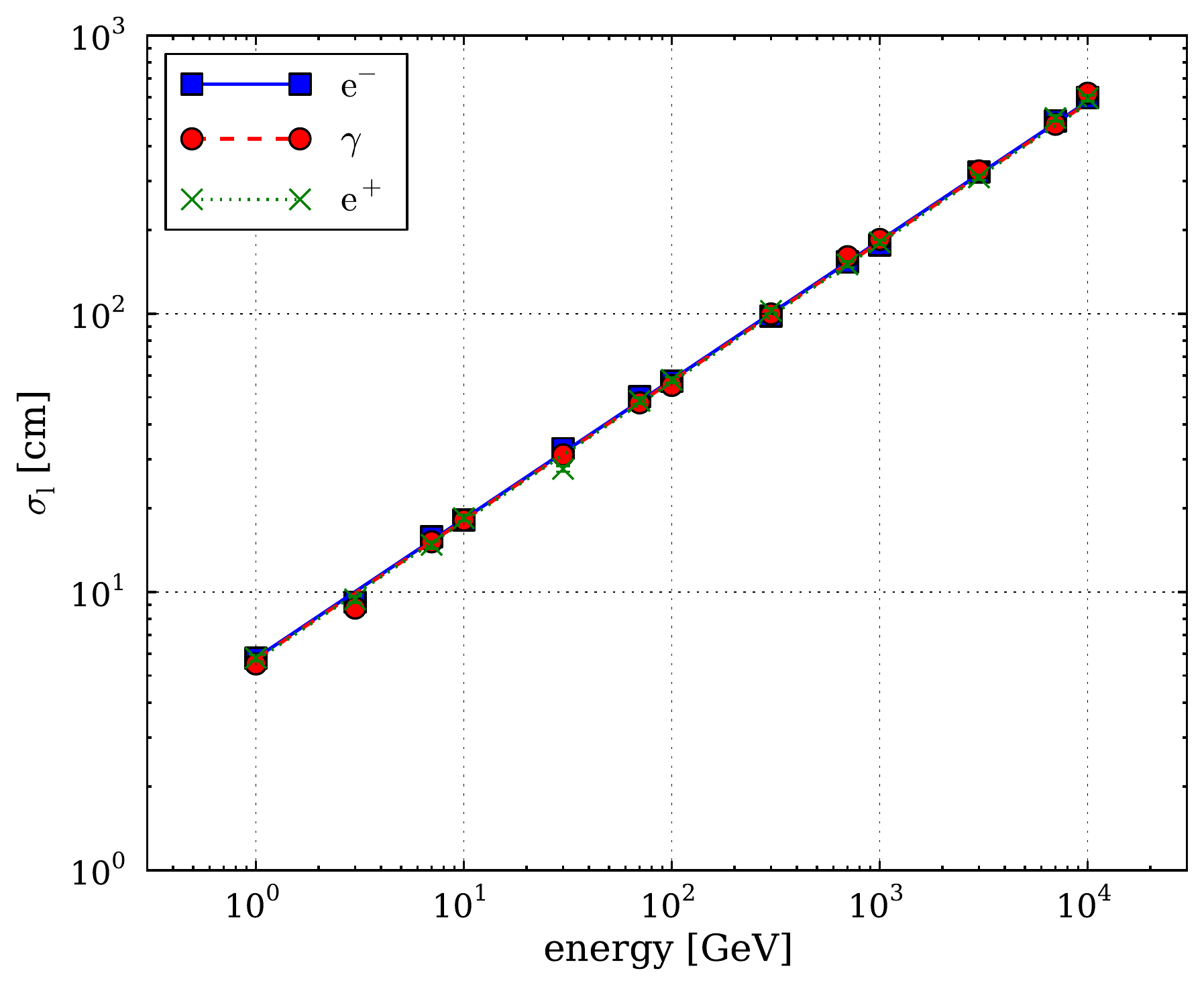}
    \caption{Total amount of Cherenkov-light-radiating track length $\hat l$, including the \tammf. \label{fig:totaltrack}
Top: distribution of $\hat l$ for $5\cdot 10^4$ simulated primary electrons of $E_{0}=1\unit{TeV}$.  A Gaussian-distribution is fit to the data. 
Middle:  $\hat l (E_0) $ as a function of  the primary energy $E_{0}$ resulting from Gaussian fits. A power-law  (Eq.(\ref{eq:totaltrack}) is  
fit to the data.
Bottom:  Standard deviation $\sigma_{\hat l} (E_0) $ Resulting from Gaussian fits. A power-law (Eq.(\ref{eq:totaltrack})) is  fit to the data.
}
\end{figure}

Figure \ref{fig:totaltrack} (top) shows as an example 
the distribution of the total  effective  track length for 
repeated simulations of a primary electron of $1 \unit{TeV} $.
The distribution can be well described by a Gaussian, which is fit to the 
data.

The mean expectation and the standard deviation from Gaussian fits
to  distributions for different primary particles and different primary energy $E_0$ are shown in figure  \ref{fig:totaltrack} (middle) and (bottom).
The data is fit with a power-law
\eqb\label{eq:totaltrack}
\hat{l}(E_{0}) = \alpha \cdot E_{0}^{\beta}  
\quad , \quad \sigma_{\hat{l}}(E_{0}) = \alpha \cdot E_{0}^{\beta} ~.
\eqe
In all fits the parameter $\beta $ is found to be consistent with $1$
at the level $10^{-5} $ indicating a very good linear relation between
the total $\hat l$ and $E_0$. Also the coefficients $\alpha $ agree within
$10^{-3} $ for different primary particles. The detailed
 results of these fits are given in table \ref{tab:results:para_l_sigma_emcascade}
in \ref{app:parares}.
As a result we obtain an energy scale parameter which relates linearly 
the 
total \cherl yield with the primary energy
\eqb
\alpha \approx 532.1 \pm 10^{-3} \, \unit{cm}\unit{GeV}^{-1} ~.
\eqe
The observed value $\alpha =  532.1 \unit{cm\,GeV^{-1}} $ is slightly larger than
 the value  $\alpha =  521 \unit{cm\,GeV^{-1}} $ in \cite{KOWALSKI}
which was also obtained in \geantfour simulations of ice. The origin of this $2\% $ difference is not  obvious. It 
could be either related  to 
difference in the versions of \geantfour but also to 
different configurations which are not given in \cite{KOWALSKI}, e.g. a slight difference in the assumed   index of refraction.

If rescaled to the density of water by the relation 
\eqb
\rho_{water} \cdot \alpha_{water} \approx  {\rho_{ice}} \cdot {\alpha_{ice}} 
\eqe
we obtain the value $\alpha_{water} \approx  484 \unit{cm\,GeV^{-1}} $.
This is significantly larger than the values  
$437 \unit{cm}\unit{GeV}^{-1}$ found in \cite{CHWPHD} for $n=1.33$ 
and $466 \unit{cm}\unit{GeV}^{-1}$ in \cite{MIRANI} for $n=1.35$.

As expected for an increasing number of particles, the size of fluctuations 
 of the total track length $\sigma_{\hat{l}} $ 
increases $\propto \sqrt{E_0} $
and  the value $\beta = 0.5 $ is fixed for the fit.
Hence,  the relative  size of
fluctuation decrease $\propto 1/\sqrt{E_0} $ with higher primary energy $E_0$
(see figure \ref{fig:totaltrack}). With the values  in
 table \ref{tab:results:para_l_sigma_emcascade}  in \ref{app:parares}  the relation 
\eqb
{\sigma_{\hat t}  \over \hat l} \approx 0.0108 \cdot \sqrt{1 \unit{GeV} \over E_0}
\eqe
is found.


\subsection{Longitudinal cascade development \label{sec:lon}}

\begin{figure}[htp]
    \centering
    \includegraphics*[width=.49\textwidth]{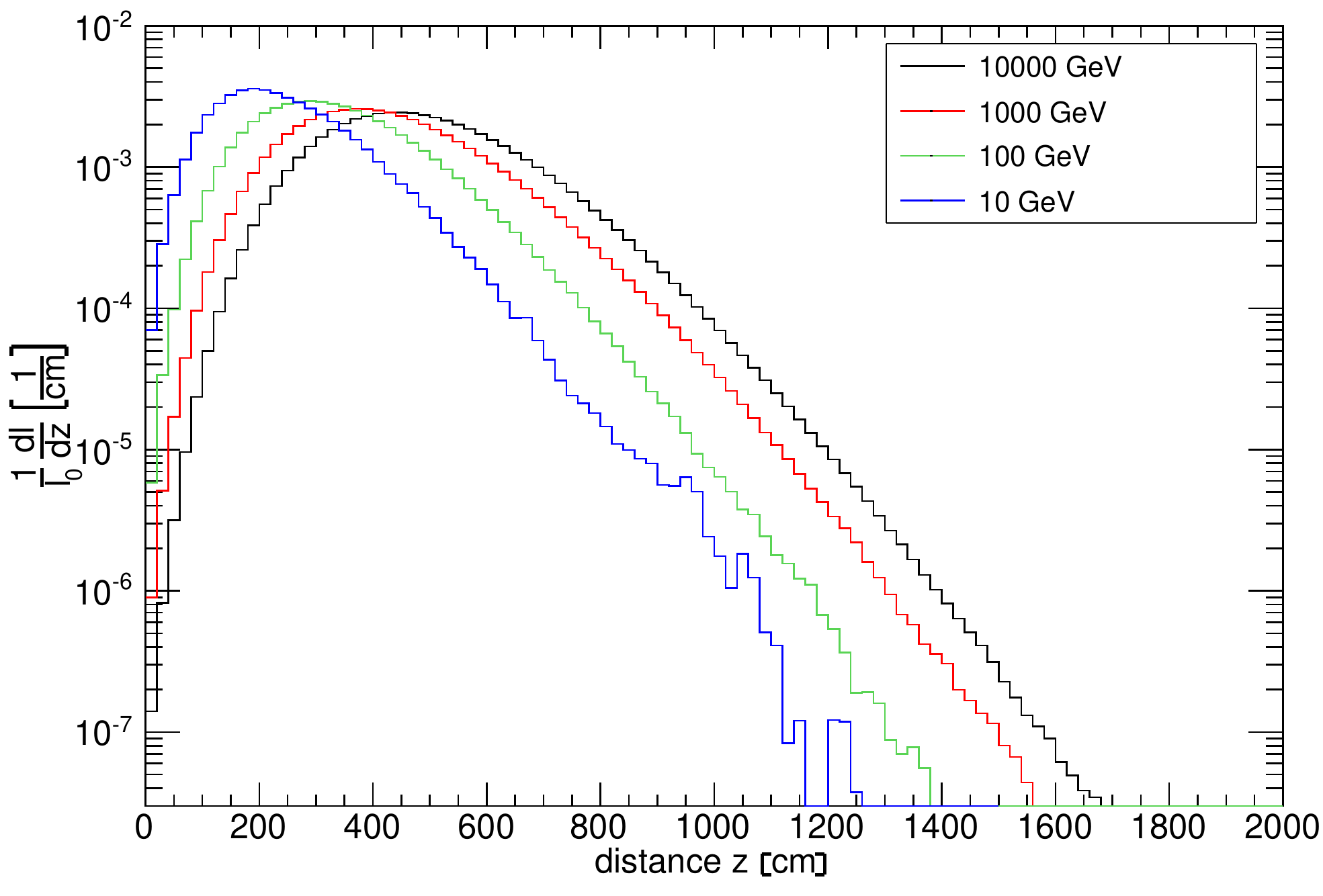}
    \includegraphics*[width=.49\textwidth]{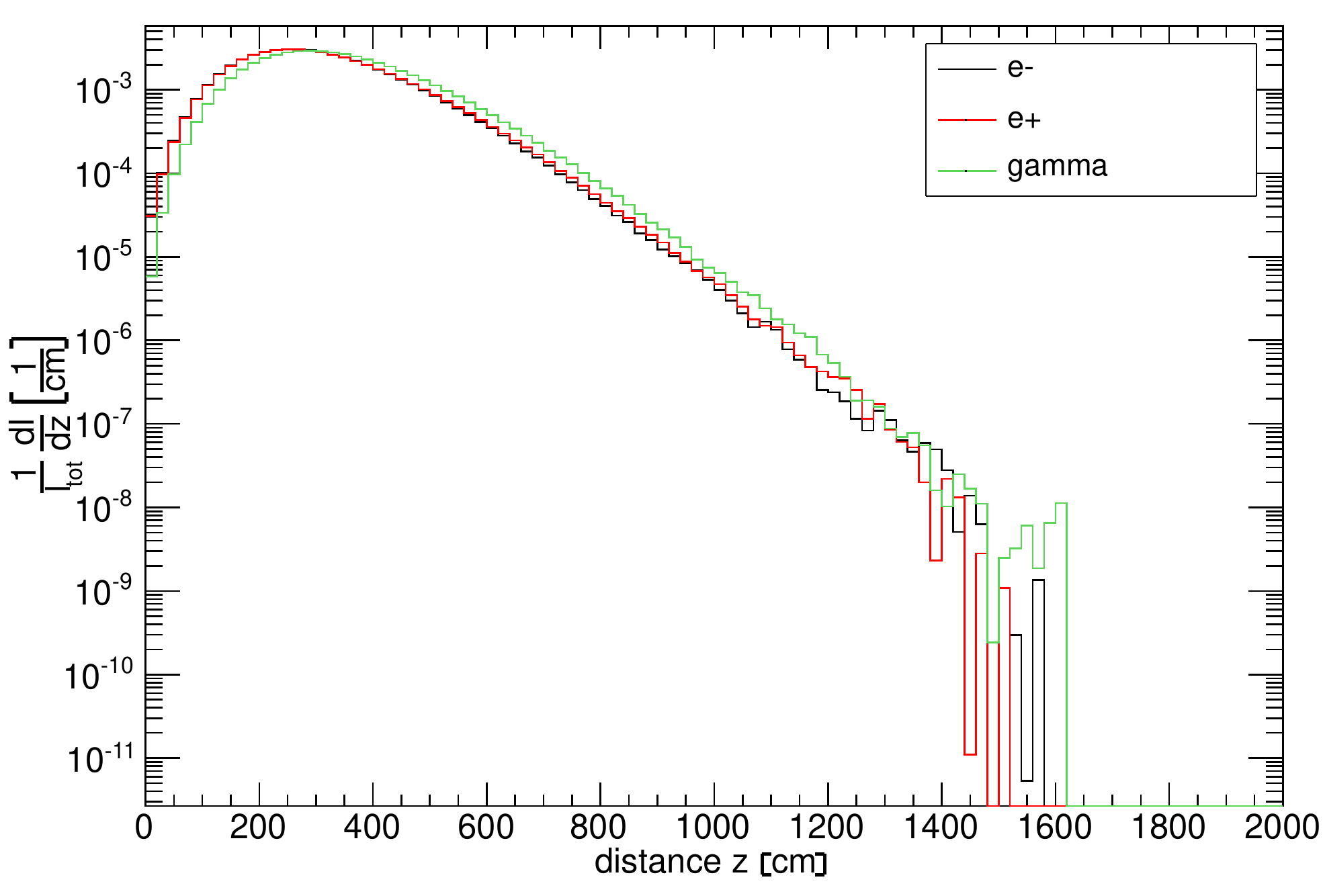}
    \caption{Longitudinal shower profiles as a function of the length $z$ along the shower axis. Shown is the  track length distribution $d\hat{l}\over dz $ relative to the total length $\hat{l}_0 $ of the cascade. The left figure
shows  the result for initial $\gamma$ and different primary energies $E_{0}$. The right figure shows the result for different primary particles $e^\pm,\gamma $ and a primary energy $E_{0}=100\unit{GeV} $.
All track segments have been  weighted with the \tammf.
     \label{fig:longdist} }
\end{figure}

The  longitudinal profiles of the  track length $d \hat l /dz  $ along
the axis of the  cascade are shown in figure \ref{fig:longdist}.
As expected from a simple Heitler model \cite{PDG,heitler1954quantum}, the depth of the shower maximum $z_{max} $ scales
logarithmically with the primary energy. 
The distributions are almost identical for $e^+$ and $e^-$. 
However, for an initial photon the depth 
 of the shower maximum is about one radiation length deeper.

\begin{figure}[htp]
    \centering
    \includegraphics*[width=.45\textwidth]{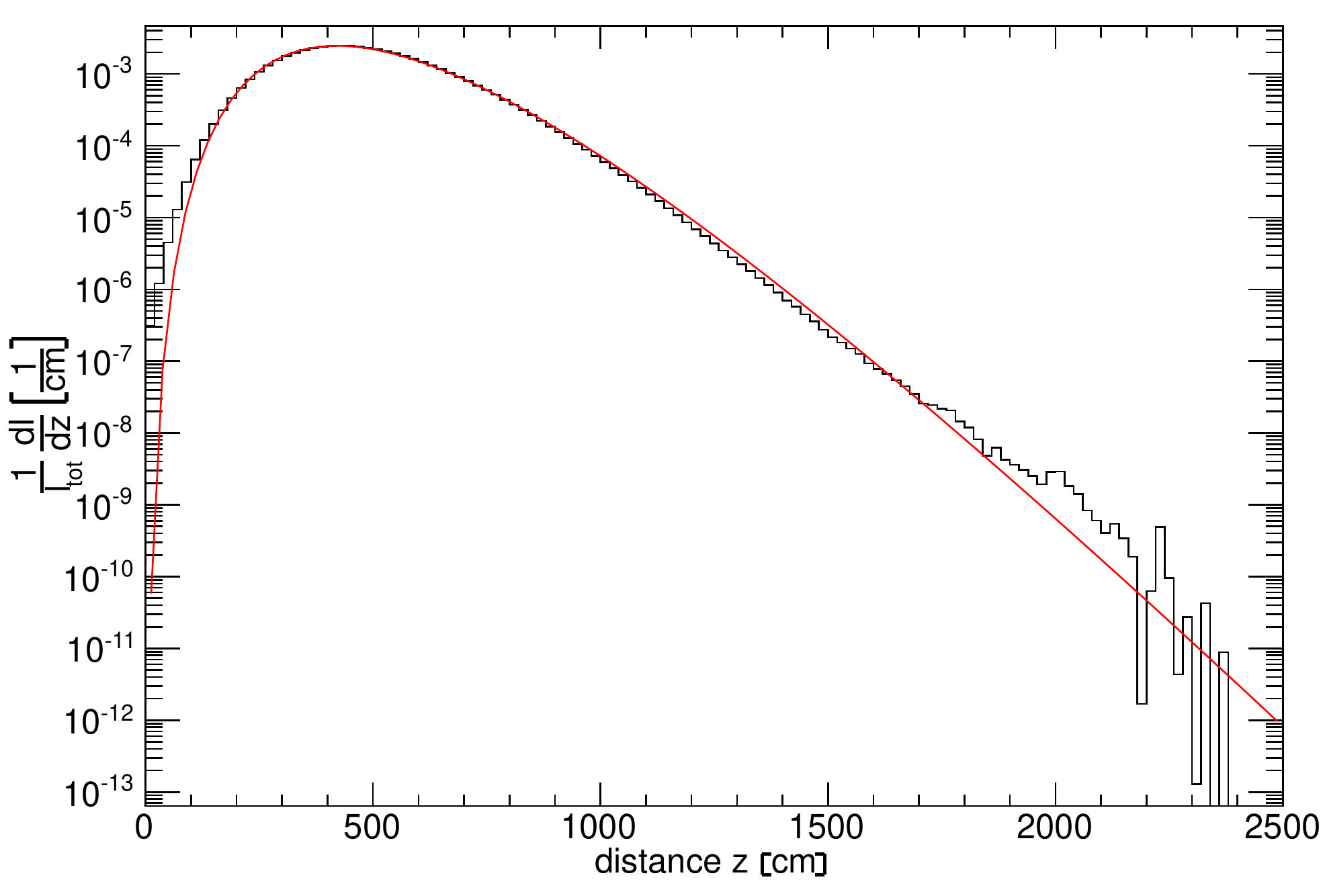}
    \caption{Example of the parameterization of the longitudinal shower profile with Eq.(\ref{eq:long:fit}) for a positron with a primary energy
    $E_{0}=10\unit{TeV}$. 
 \label{fig:longdistfit}}
\end{figure}

The longitudinal shower profile can be parameterized with a gamma distribution
\eqb \label{eq:long:fit}
    \left.\widehat{l}\right.^{\,-1}_{\mathrm{tot}}\cdot\frac{\ud\widehat{l}}{\ud t} = b\,\frac{\left(bt\right)^{a-1}\mathrm{e}^{-bt}}{\Gamma\left(a\right)} ~.
\eqe
Here, $t$ is the shower depth $ t  \equiv z / X_0  $,
 $a$ and $b$ are characteristic dimensionless constants  \cite{PDG}.
An example fit is shown in figure \ref{fig:longdistfit}.
The results of all fits for different $E_0$ are given in table \ref{tab:reslon} in \ref{app:parares}.

\begin{figure}[htp]
    \centering
    \includegraphics*[width=.65\textwidth]{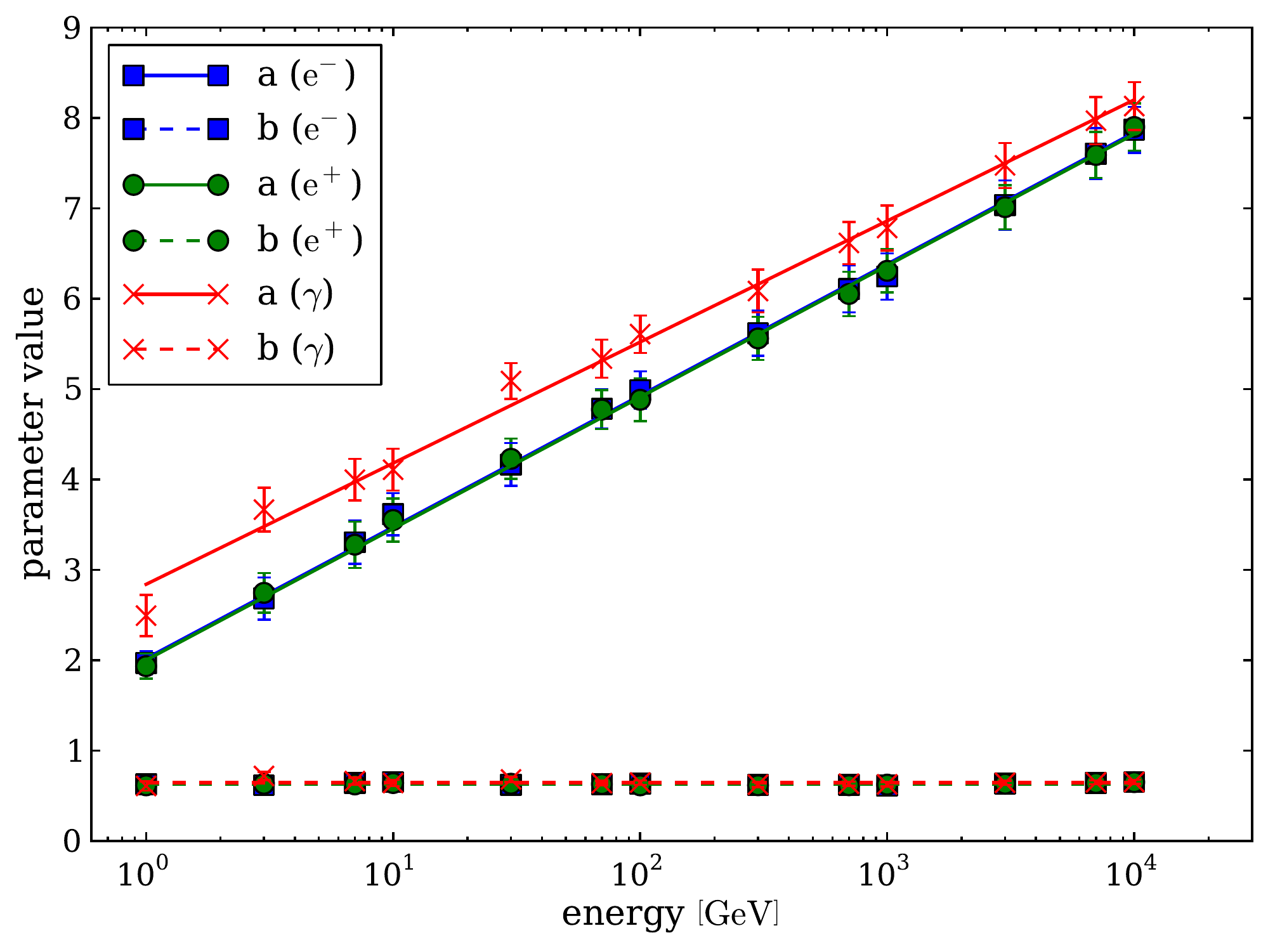}
    \caption{Fit parameters of the longitudinal profile $a$, $b$  versus initial energy.
 \label{fig:longdistpar}}
\end{figure}

The energy dependence of the fit parameters
 $a$ and $b$ is shown in figure \ref{fig:longdistpar}.
$b$ is found constant and does not depend on the particle type, while
the parameter  $a$ can be described  with an logarithmic increase
\begin{equation}
    a = \alpha+\beta \cdot  \log_{10} \lbra {E_0 \over 1\unit{GeV} } \rbra  ~.
\end{equation}
It  is slightly larger for $\gamma $ than for $e^\pm$.
The parameterization results for  $b$ and $\alpha $, $\beta $
are given in table \ref{tab:emlong::energypar} in \ref{app:parares}.

\begin{figure}[htp]
    \centering
    \includegraphics*[width=.95\textwidth]{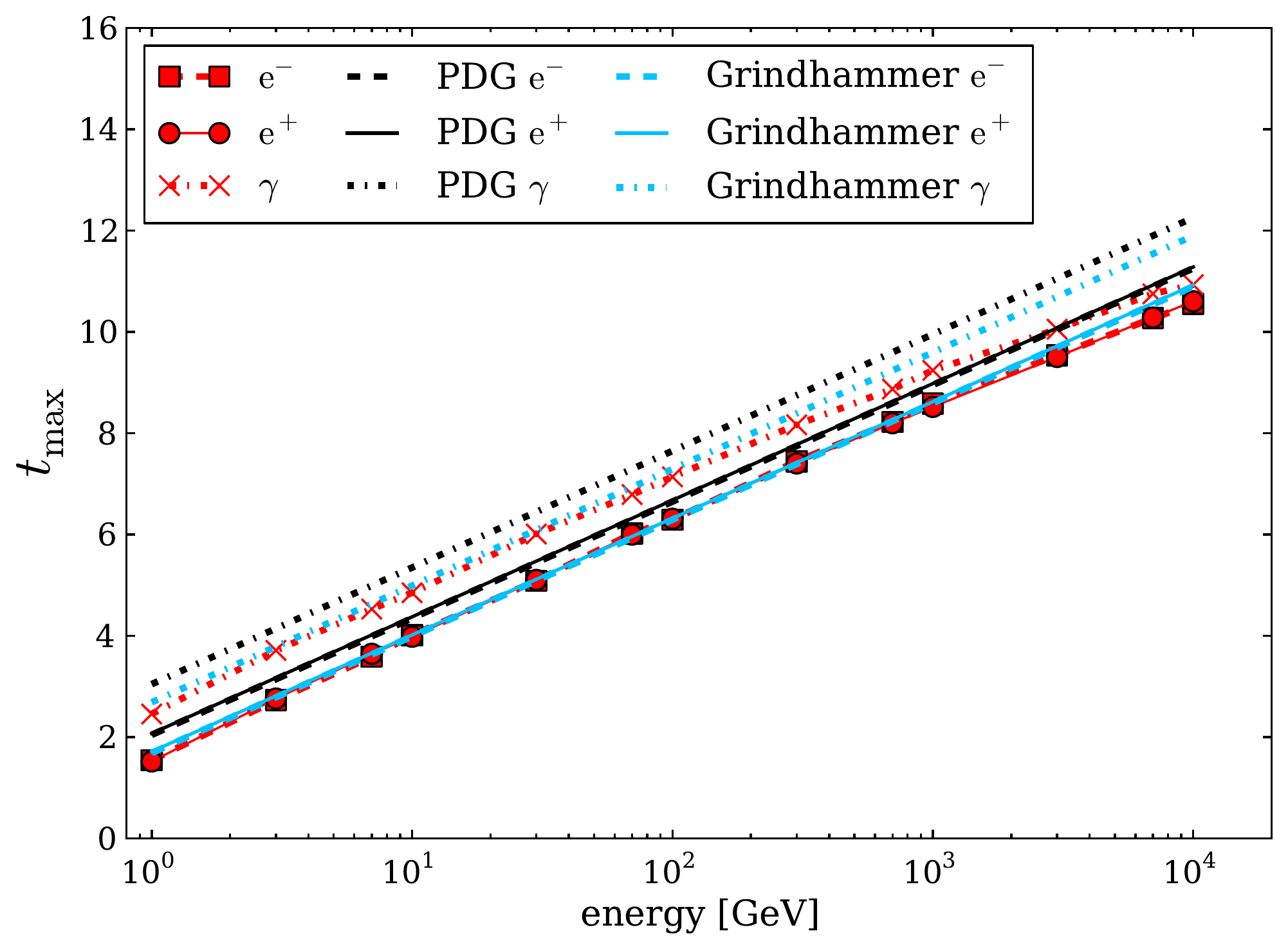}
    \caption{Maximum of the longitudinal Cherenkov-radiating track-length profile as a function of the initial energy. The markers represent the calculated values for $t_{\mathrm{max}}$ and solid lines are for visual guidance. The dotted/dashed lines show Eq.(\ref{eq:em:tmax_pdg}) with the parameters from \cite{PDG} (PDG)/\cite{Grindhammer}(Grindhammer) respectively.
    \label{fig:em:tmax}} 
\end{figure}


Figure \ref{fig:em:tmax} shows the shower maximum of the Cherenkov-radiating 
track length as a function of the initial energy. It has been
 calculated from the longitudinal shower profiles with the formula
\begin{equation}
    t_{\mathrm{max}} = \frac{a-1}{b}.
\end{equation}
Additionally shown is the  maximum of the longitudinal energy deposition 
based on  the parameterization in \cite{PDG} 
\begin{equation}\label{eq:em:tmax_pdg}
    t_{\mathrm{max}} = \ln y + C_{\mathrm{j}},\qquad \mathrm{j}=\mathrm{e},\gamma,
\end{equation}
with  $y=E_{0}/E_{\mathrm{crit}}$.
In \cite{PDG} for electron- or positron-induced cascades the values $C_{\mathrm{e}}^{\mathrm{PDG}}=-0.5$ and for photon-induced cascades  $C_{\gamma}^{\mathrm{PDG}}=+0.5$
are given. These values are based on simulations with \textit{EGS4} up to an energy of $100$\, GeV for nuclei heavier than carbon. Up to that energy the slope agrees with our result  but our values are offset by about $-0.5$. Above
$100 \unit{GeV} $ we also deviate in slope.

In contrast,   \cite{Grindhammer} gives 
the value  $C_{\mathrm{e}}^{\mathrm{Gr.}}=-0.858$ 
 for electron- or positron-induced cascades.
This parameter was also obtained from fits to the longitudinal energy deposition 
profiles for elements ranging from carbon to uranium at energies 
from $1\unit{GeV}$ to $100\unit{GeV}$. The simulations were performed with
\textit{Geant3}.
The value $C_{\mathrm{\gamma}}^{\mathrm{Gr.}}$ is not explicitly stated.
Assuming that the difference of the maxima of photon- and electron-induced 
cascades is about one radiation length we obtain $C_{\mathrm{\gamma}}^{\mathrm{Gr.}} \approx  C_{\mathrm{e}}^{\mathrm{Gr.}}  +1 =  +0.142$.

Here, the simulations are performed with \geantfour and ice is used as the detector material.
It can be seen that our results agree much better  with 
\cite{Grindhammer}.
Nevertheless, deviations appear  for larger energies, where the here obtained
 energy dependence increases  less than logarithmically.
This effect is stronger for photon-induced cascades than for electron-induced cascades.


\subsection{Angular distribution of tracks \label{sec:angtrack}}

\begin{figure}[htp]
    \centering
    \includegraphics*[width=.49\textwidth]{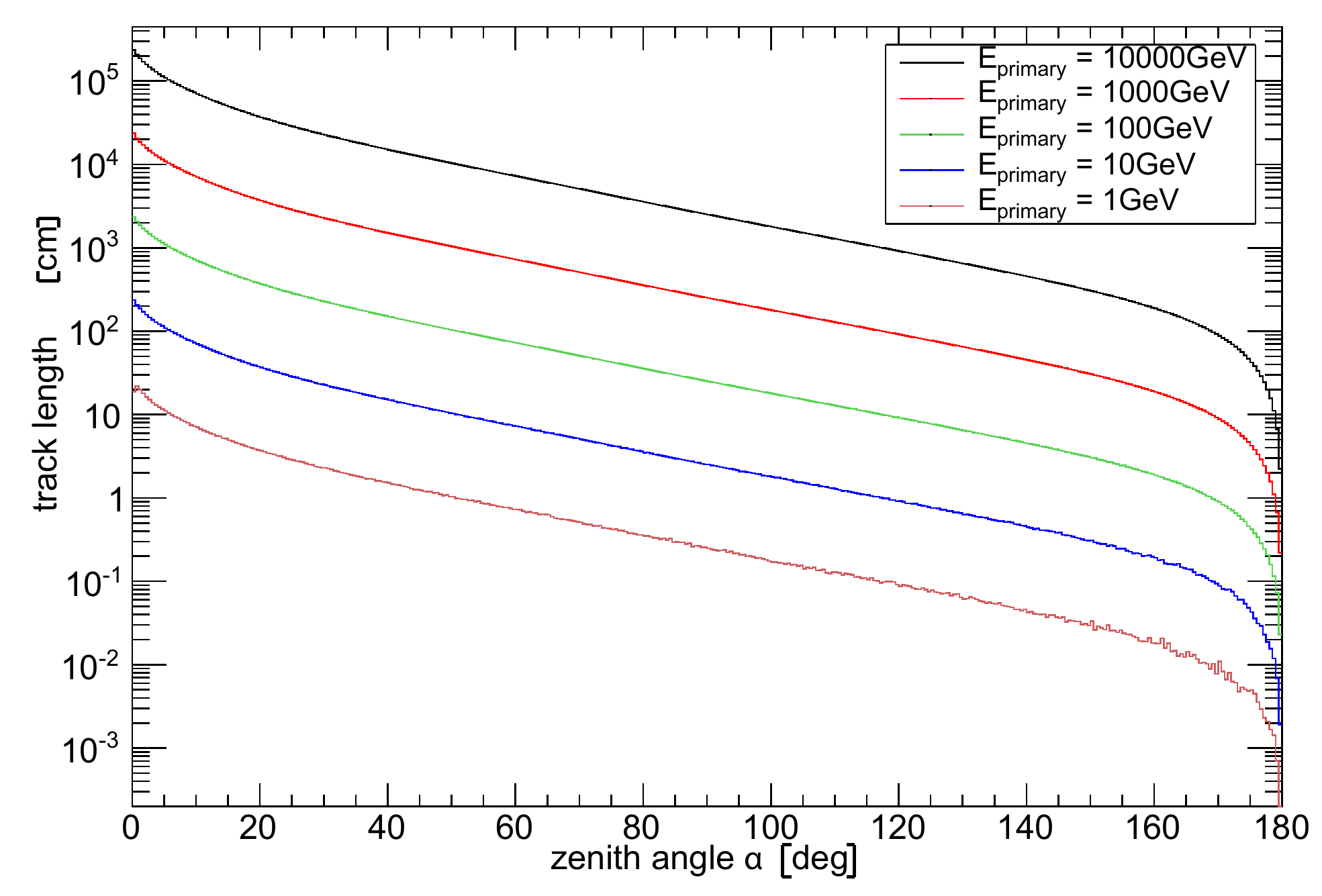}
    \includegraphics*[width=.49\textwidth]{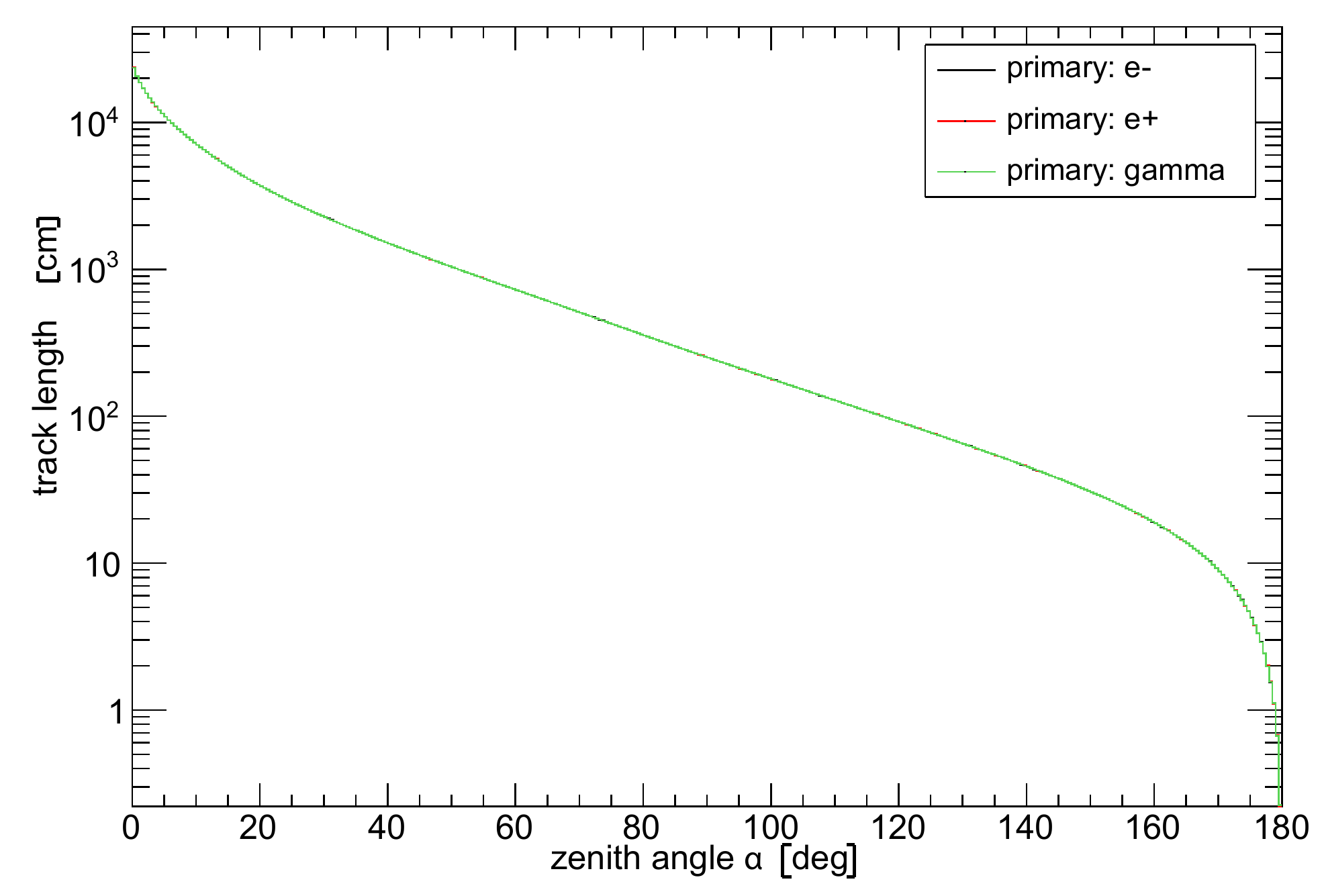}
    \caption{Distribution of the effective track length $\hat l $  per shower as a
        function of the inclination angle $\alpha$ of the primary particle's direction. The left figure shows the result
        for a primary positron and different $E_{0}$ and the right figure the 
results for
        $E_{0}=1\unit{TeV}$ and different primary particles. 
     \label{fig:alpha}}
\end{figure}

\begin{figure}[htp]
    \centering
    \includegraphics*[width=.65\textwidth]{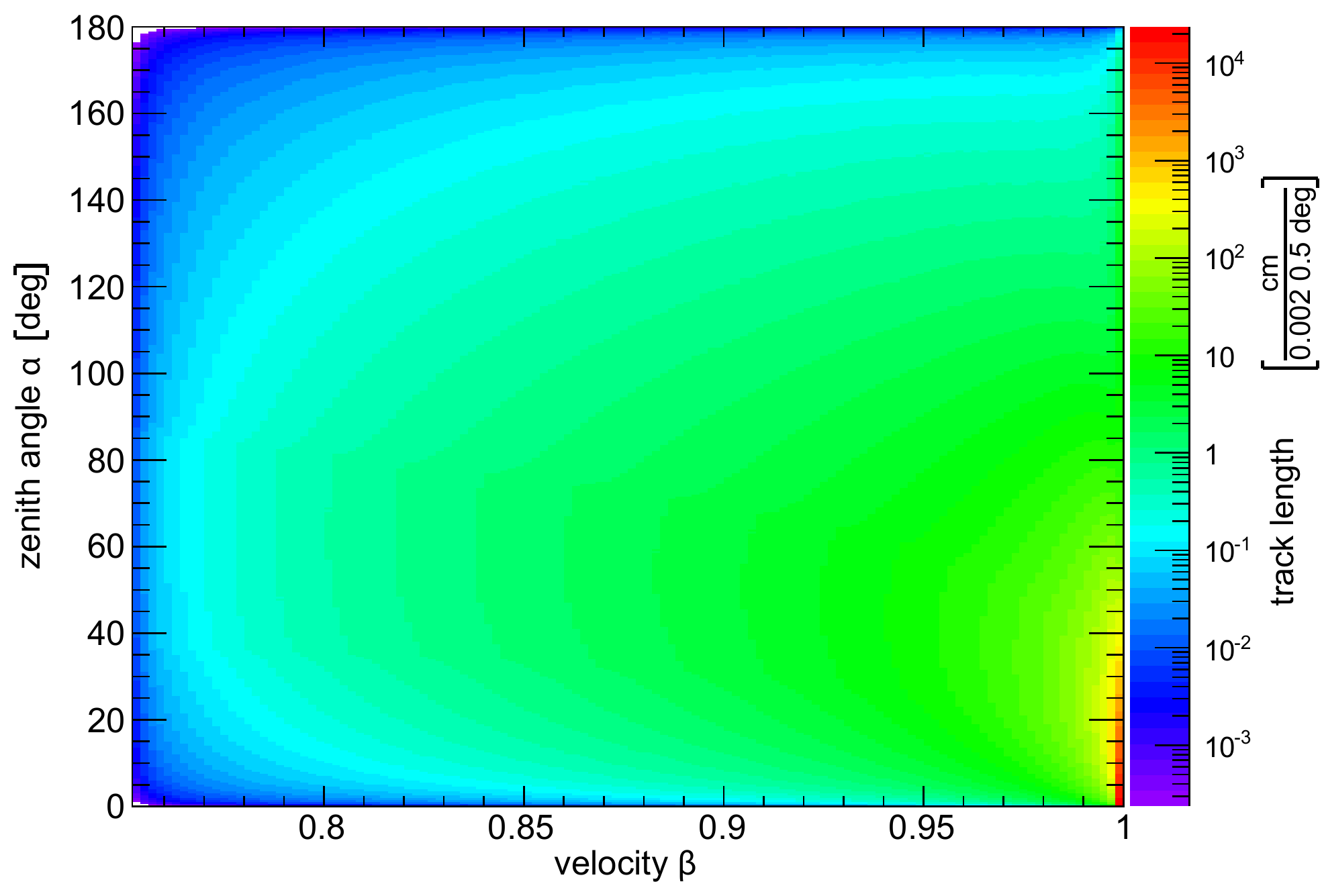}
    \caption{Density distribution of the effective track length $\hat l $ 
versus the inclination angle $\alpha$ and Lorentz factor $\beta$ for a $1\unit{TeV}$ shower. The vertical color codes corresponds to the 
histogrammed length $\hat l$ per shower.
     \label{fig:alphabeta}}
\end{figure}

Relevant for the angular distributions of \cherphos is the angular distribution of secondary tracks in the cascade and their velocity
$ \frac{d^2 \hat{l}}{d \alpha \, d\beta} $. Here, $\alpha $ is the polar angle of the track with respect to the $z$ axis and $\beta $ the Lorentz-factor.

The distribution $\frac{d \hat{l}}{d \alpha} $ is shown in figure \ref{fig:alpha} and the distribution $ \frac{d^2 \hat{l}}{d \alpha \, d\beta} $ in figure \ref{fig:alphabeta}. 

Most particles are produced in forward direction with a velocity $\beta $
close to $1$. 
It can be seen that the normalization   of the $\alpha $-distribution changes
with energy but the shape does not. The angular distribution does not change  for different primary particles.

\subsection{Angular distribution of \cherl \label{sec:angular}}

\begin{figure}[htp]
    \centering
    \includegraphics*[width=.49\textwidth]{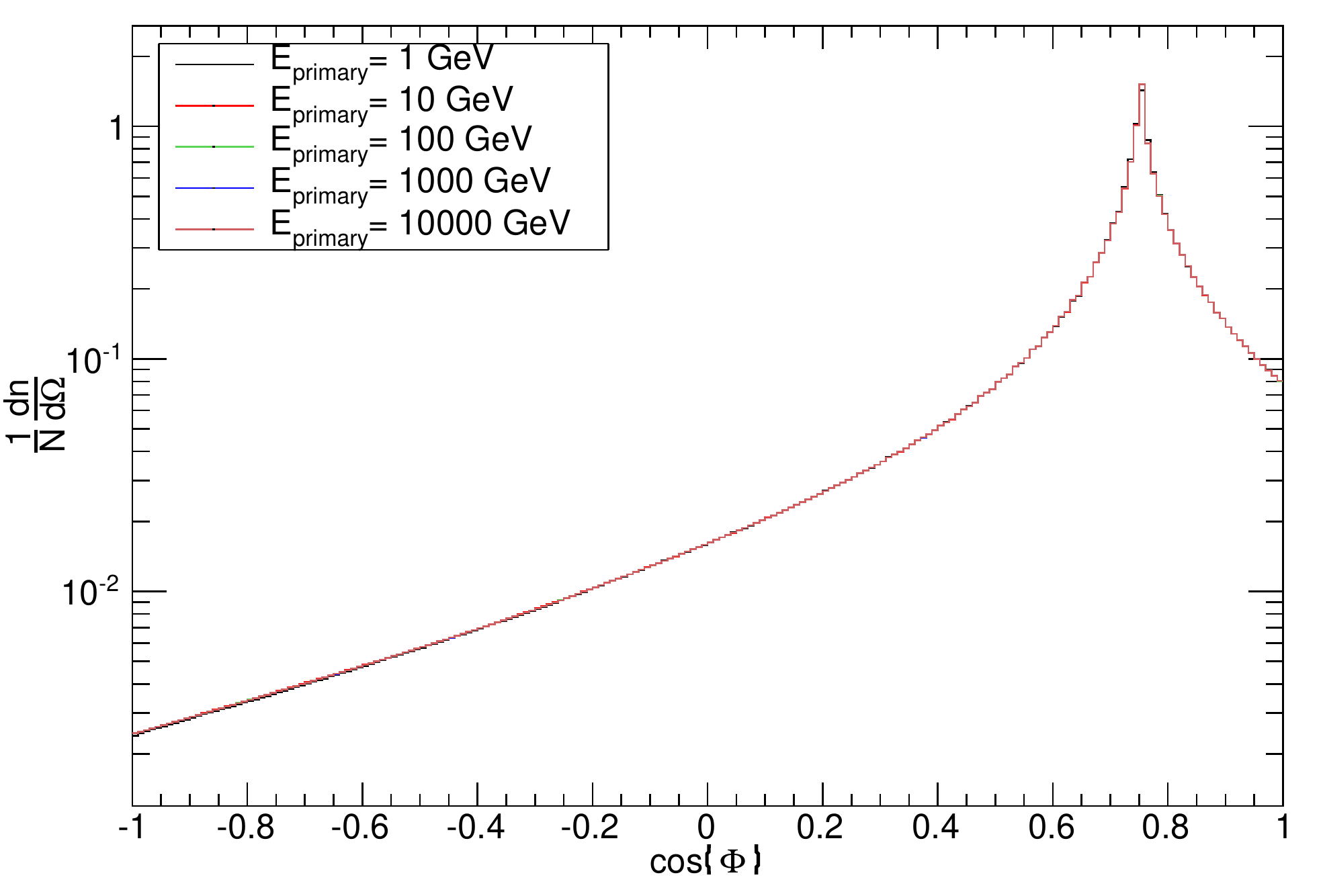}
    \includegraphics*[width=.49\textwidth]{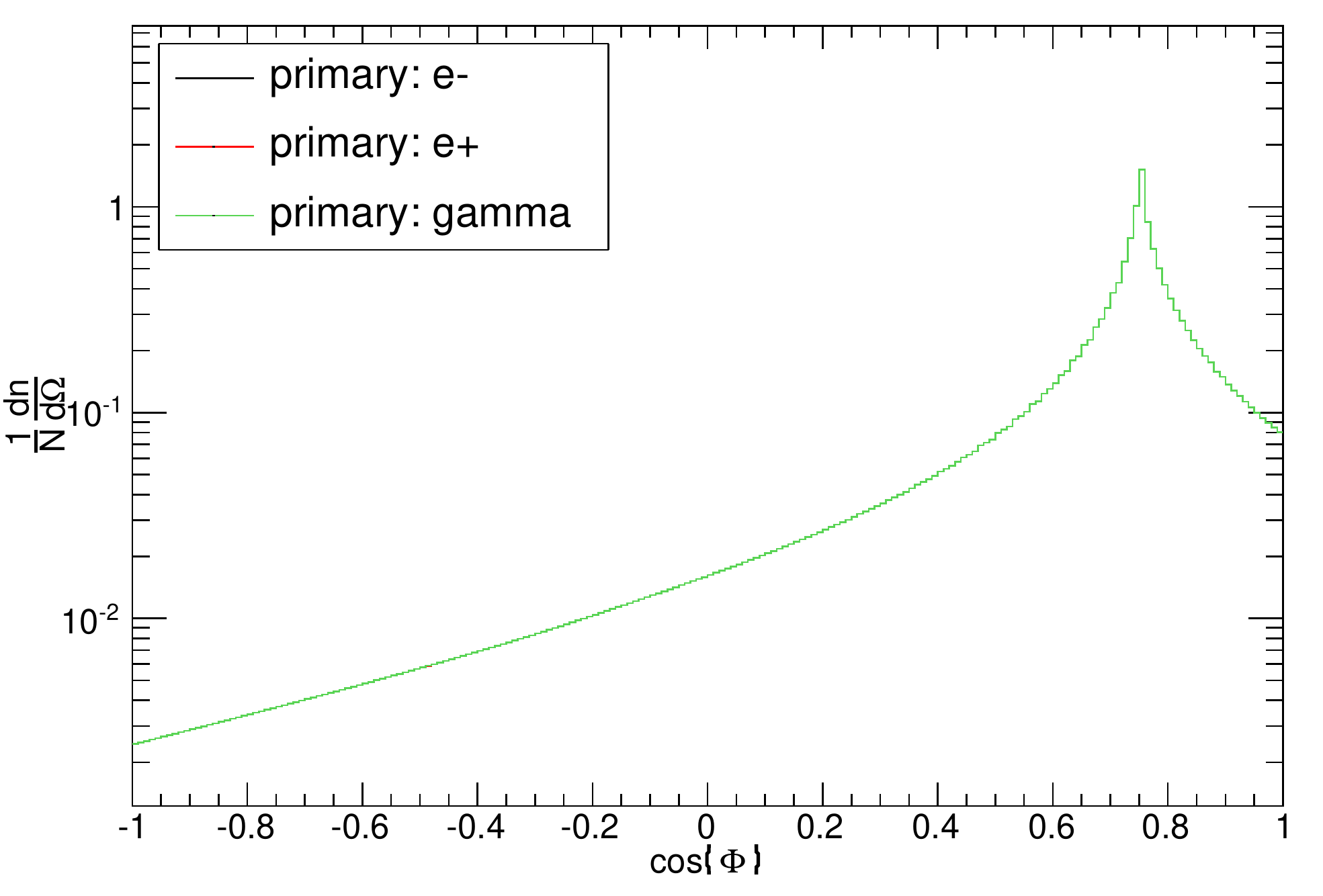}
    \caption{The angular distribution of  Cherenkov photons for different
        $E_{0}$ (left) and different primary particles (right).
        Shown are the normalized angular distributions per photon and steradian.
     \label{fig:angulardist}}
\end{figure}

The angular distribution of \cherphos  is calculated with the method
described in \cite{NAKED}. Figure \ref{fig:angulardist} shows example
 distributions $\frac{d \hat{l}}{d \Phi} $ versus the zenithal angle $\Phi $
with the z-axis\footnote{Note that the definition of $\Phi $ is 
identical to the previously defined angle $\alpha $. However we use a different symbol to indicate the difference of photons and tracks.}.
The normalization of the distribution corresponds to the track length that produces an 
equivalent total \cher light yield.

A broad distribution with a clearly pronounced 
\cher peak is visible. As expected from the results in section \ref{sec:angtrack},
the shape of the distribution is unchanged   
for different primary energies and different primary particles.

\begin{figure}[htp]
    \centering
    \includegraphics*[width=.45\textwidth]{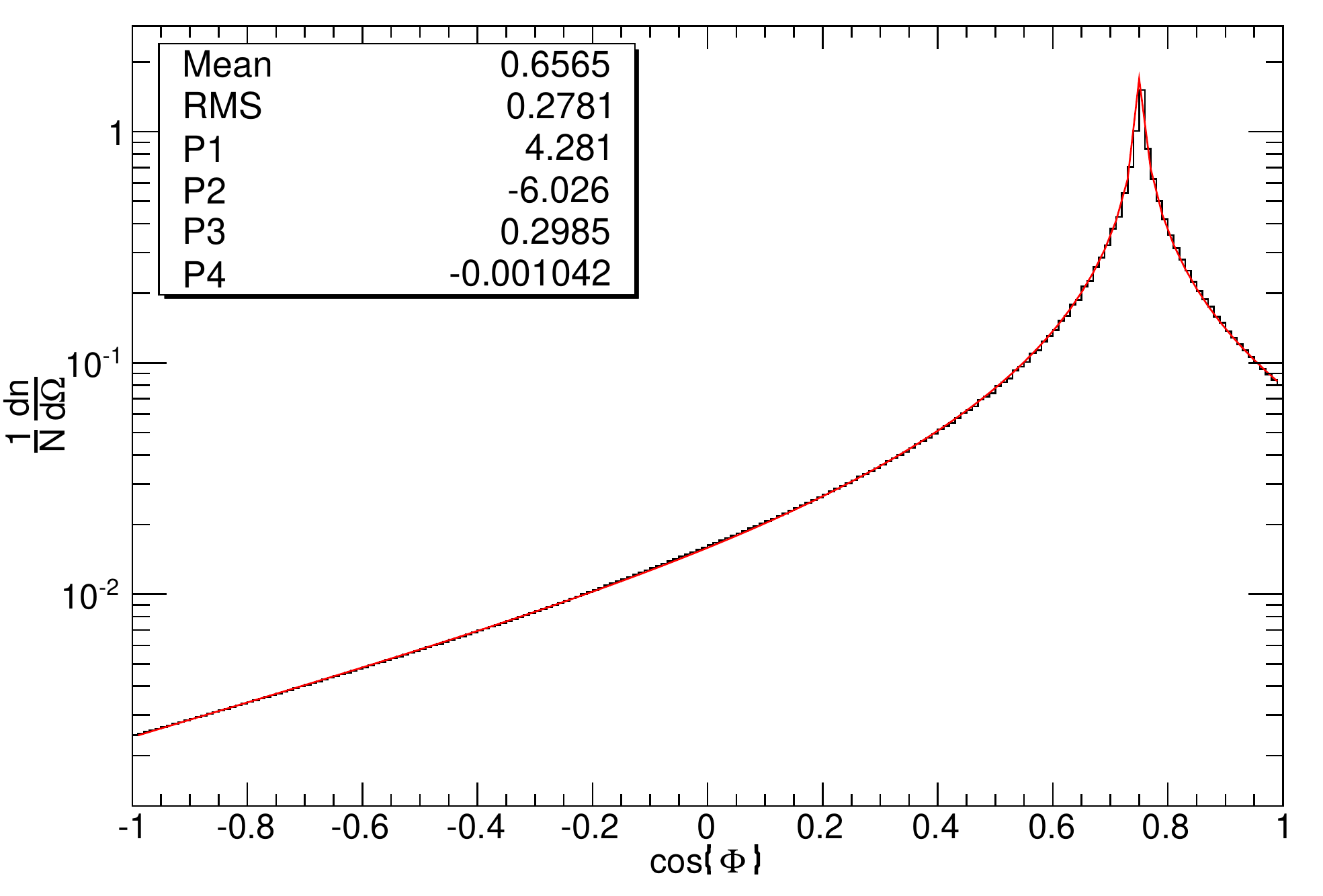}
    \caption{Example of the parameterization of the angular distribution for a 
    $E_0=100\unit{GeV}$ positron. The parameters $P1$ to $P4$ correspond to the parameters $a$ to $d$ in Eq.(\ref{eq:angpar:sym}).
     \label{fig:angulardistfits}}
\end{figure}

The angular distributions are parameterized with a simple function
\eqb
    \label{eq:angpar:sym}
    \frac{dn}{d\Omega}= a e^{b\left|x-\cos\Theta_{c,0}\right|^{c}}+d ~.
\eqe
A typical fit is shown in figure \ref{fig:angulardistfits}.

The fit parameters for different energies are given in table \ref{tab:angcher}
in \ref{app:parares}. 
They are found to be very similar and constant with energy. 
We conclude that the angular distribution of \cherphos can be described
with the above formula and the  averaged parameters
given in table \ref{tab:average:angcher} in \ref{app:parares}.

\subsection{Fluctuations in azimuth \label{sec:azi}}

\begin{figure}[htp]
    \centering
    \includegraphics*[width=.49\textwidth]{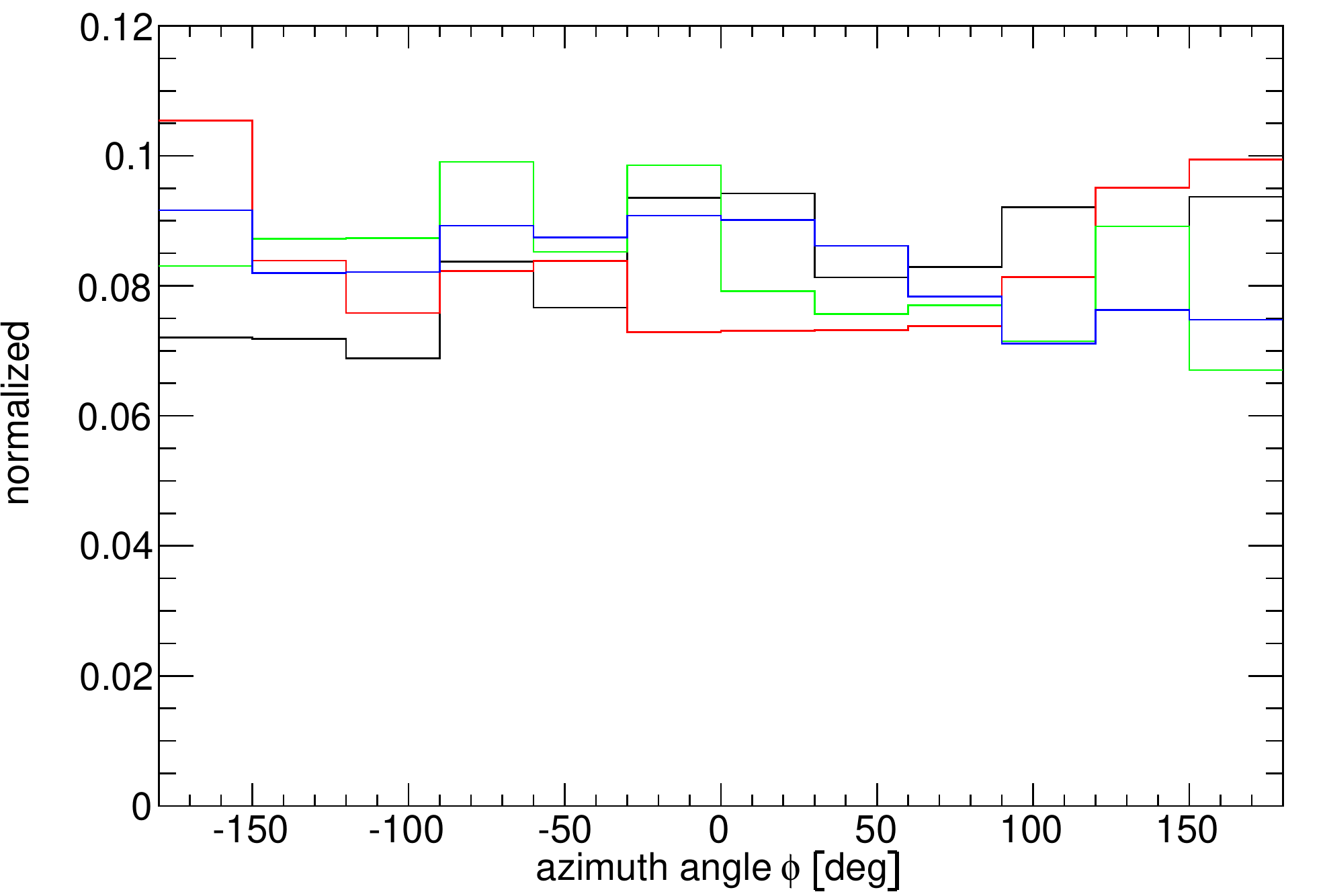}
    \includegraphics*[width=.49\textwidth]{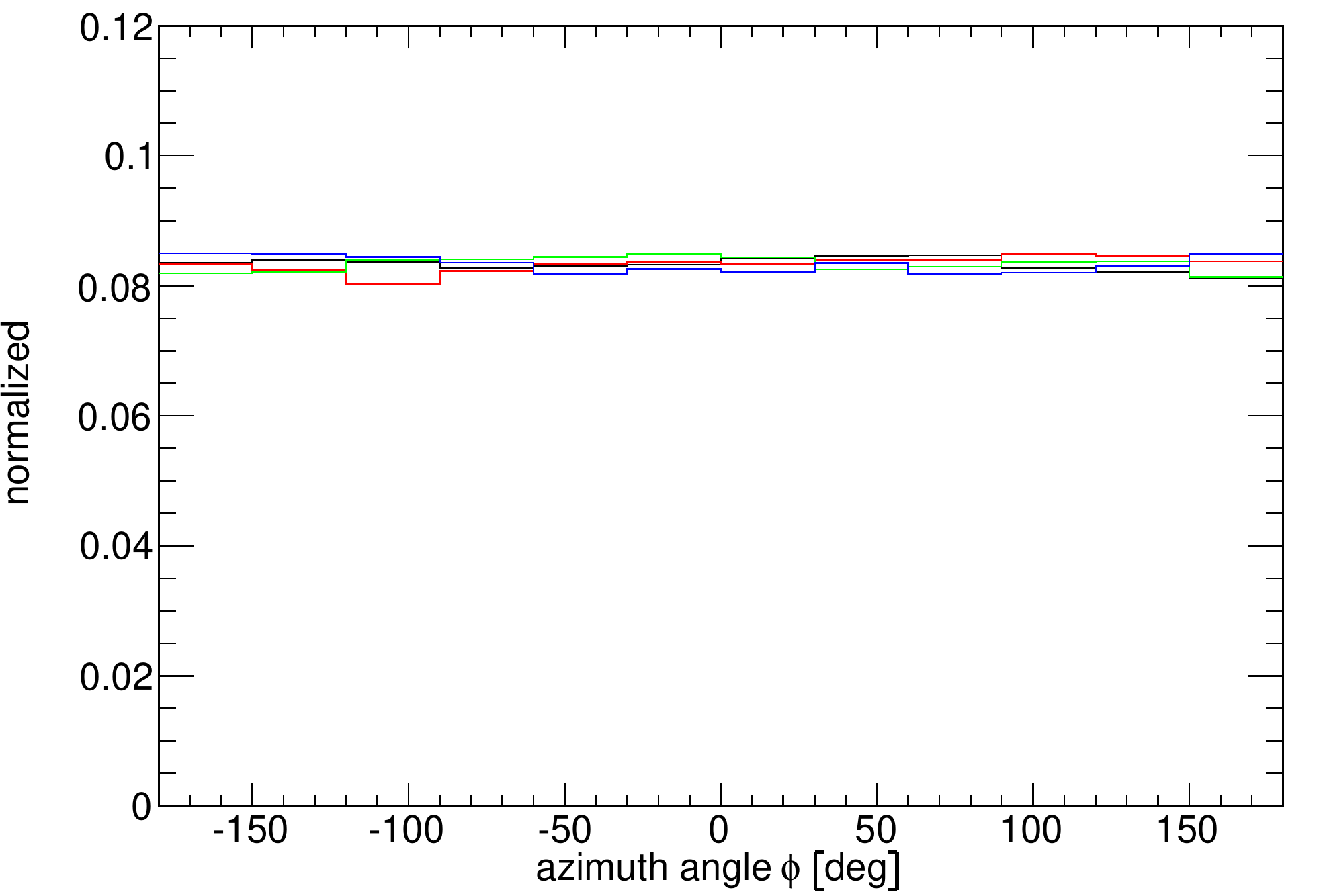}
    \caption{Example of the azimuthal distribution of four individual showers of electrons with
    a primary energies $E_0=10\unit{GeV}$ (left) and $E_0=1\unit{TeV}$
    (right). The relative effective track length per $30^{\circ}$-interval is shown.
     \label{fig:phidist}}
\end{figure}

An important prerequisite for the here used 
calculation of the angular distribution
 of \cherphos  is the assumed symmetry in 
azimuthal angle $\phi $ of  the distribution of track
directions\footnote{Note, that the azimuthal angle $\phi $ is different from the earlier defined angle $\Phi $ (see section \ref{sec:angular}) which was the zenithal emission  angle of  \cherphos.} in the plane around the shower axis \cite{NAKED}.
As an example, figure \ref{fig:phidist} shows azimuthal distributions of $\hat l$  for four individual showers, each for  two 
different energies. Differences  originate from fluctuations 
in the shower development.
Correspondingly, the  relative size of fluctuations strongly decreases for
larger particle numbers in higher energy showers and the distribution becomes almost flat in azimuth.

\begin{figure}[htp]
    \centering
    \includegraphics*[width=.49\textwidth]{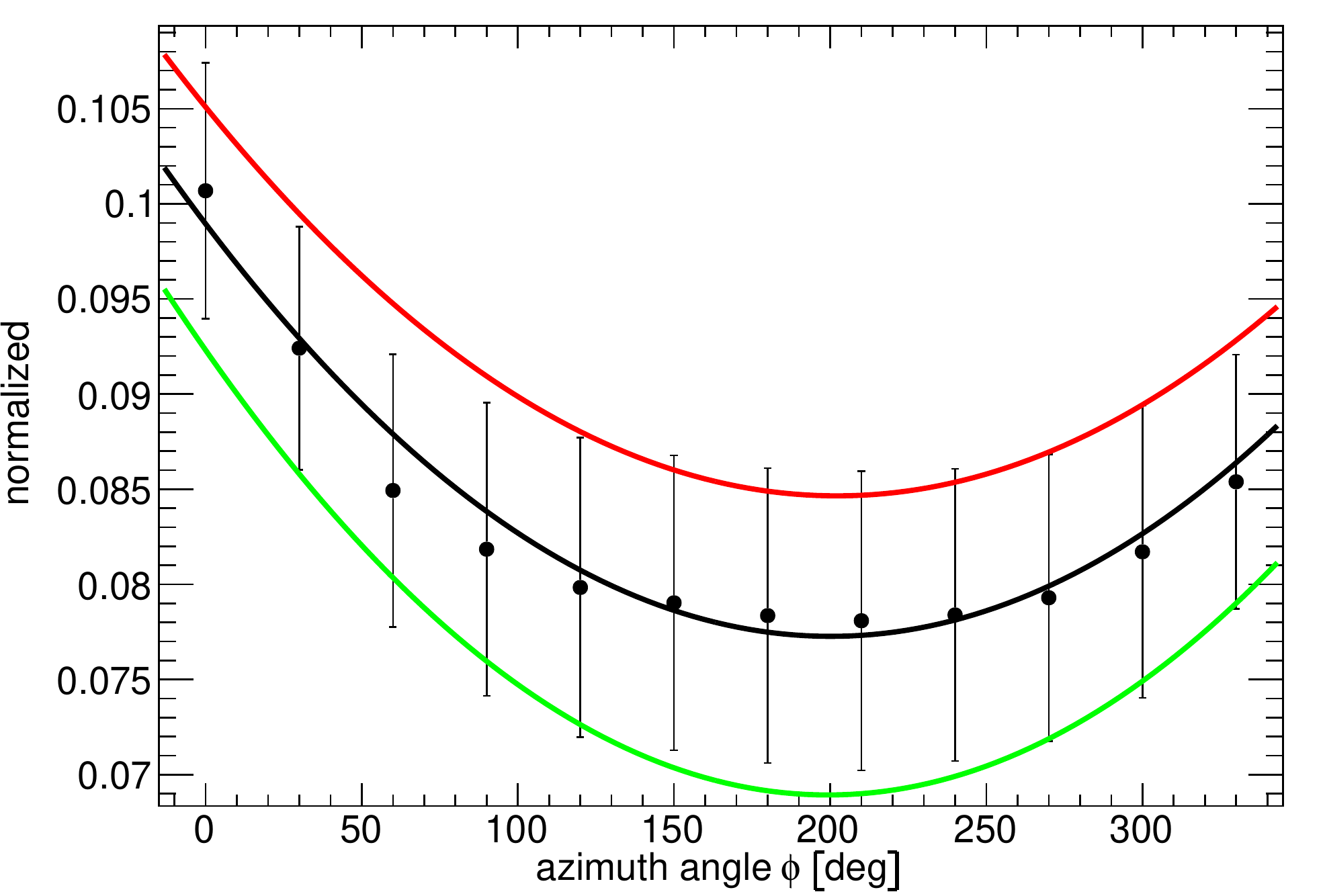}
    \includegraphics*[width=.49\textwidth]{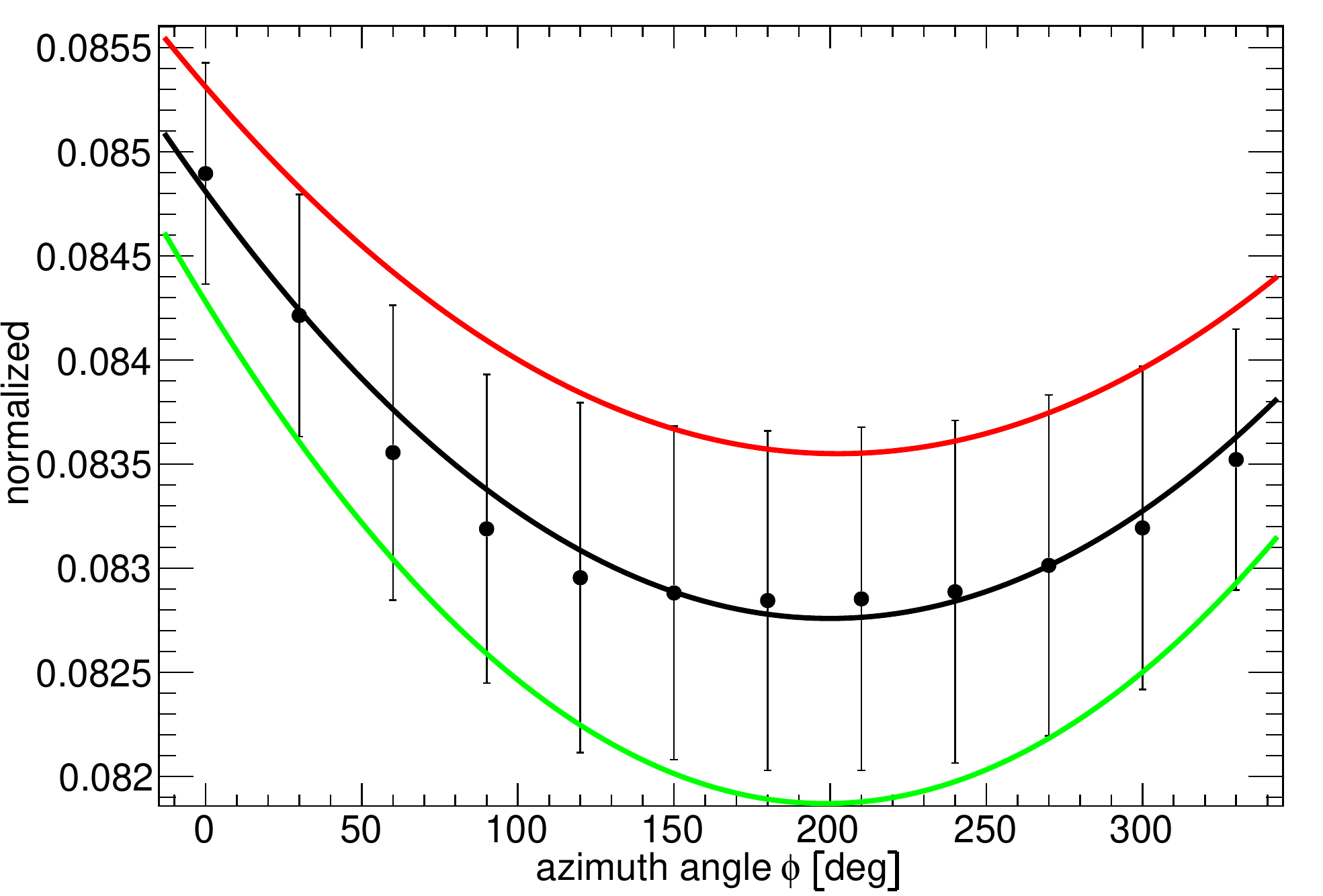} \newline
    \includegraphics*[width=.49\textwidth]{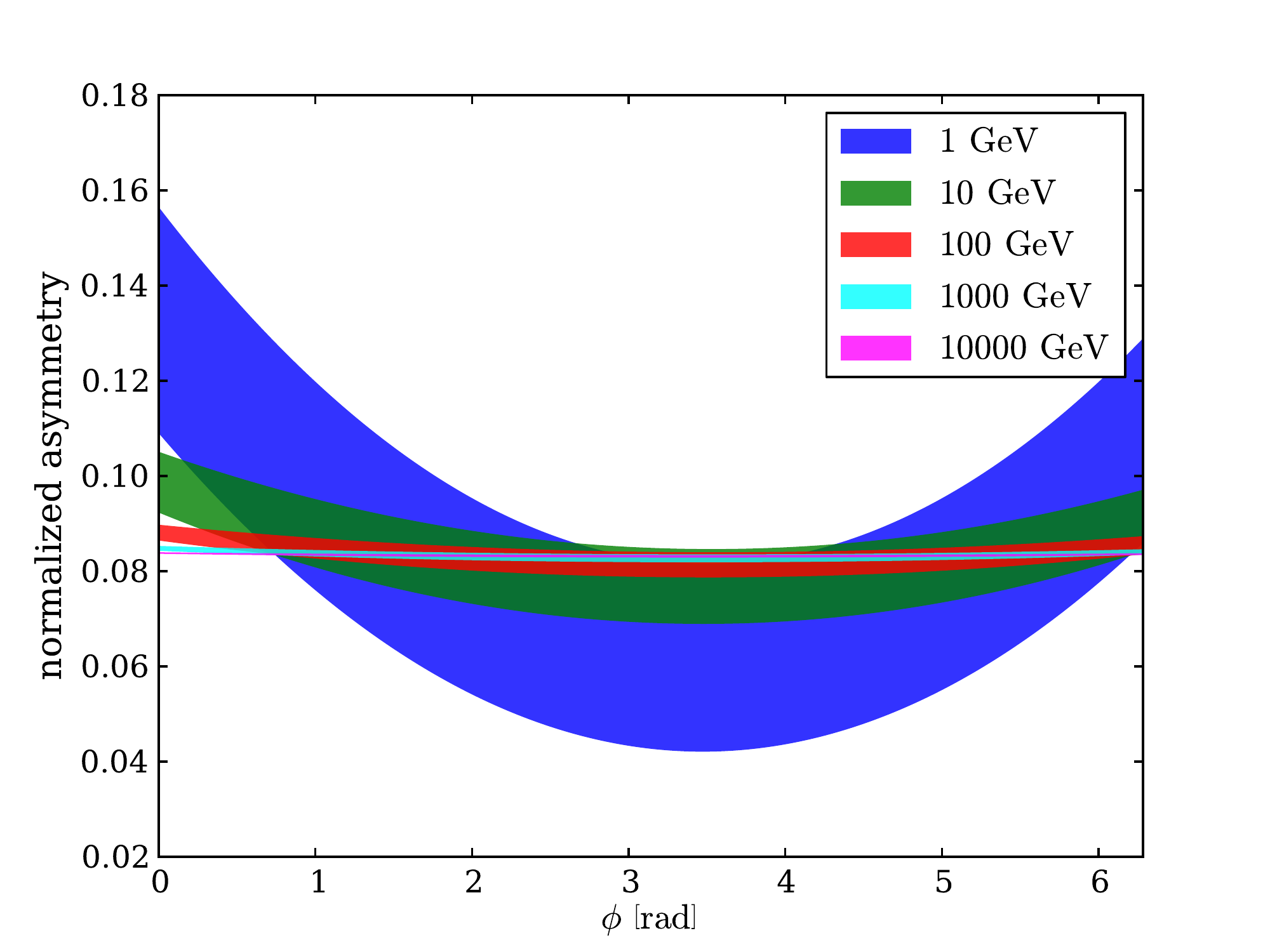}
    \includegraphics*[width=.49\textwidth]{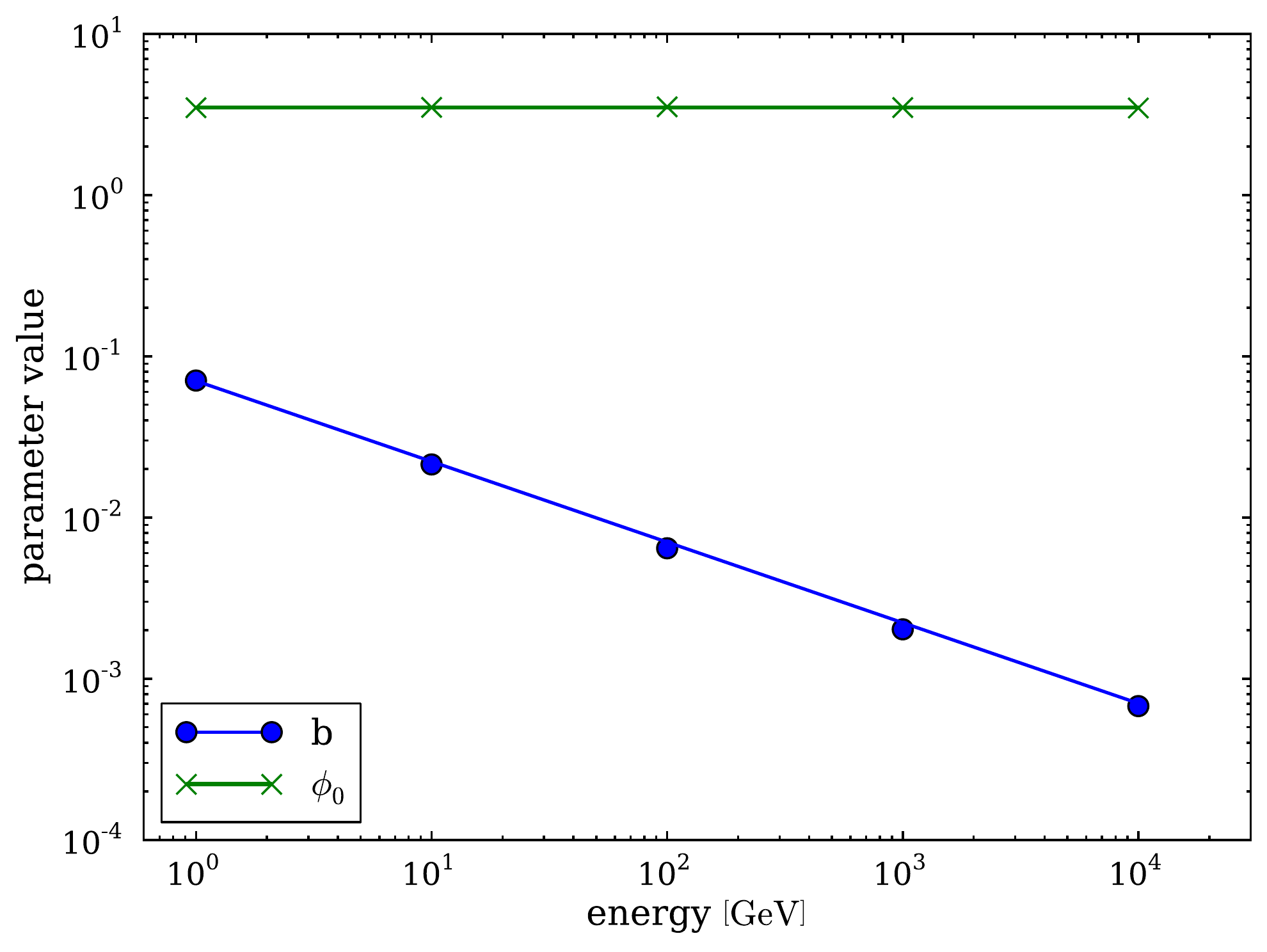}
    \caption{Fitted shower asymmetry of electron-induced showers 
for different  primary energies:  $E_0=10\unit{GeV}$ (top left) and 
$E_0=1\unit{TeV}$ (top right).
   The dots correspond to the mean expectation, if each shower is  aligned before averaging as described in the text. The shown error bars represent the 
standard deviation of  the individual bins. 
A parabola Eq.(\ref{eq:azimut:param}) was fit to the 
distributions. The black curve represents the fit to the data and 
the red and green curves fits, where  each bin content was increased or reduced 
by one  standard deviation. The bottom left figure shows the range of the fitted
uncertainties for different primary energies, and the bottom right figure
the results of the parameters in Eq.(\ref{eq:azimut:param})
versus the primary energy.
     \label{fig:phiasymmetry} }
\end{figure}

The  effect of this asymmetry can be quantified by aligning 
all simulated showers in azimuth in order to account for their 
random orientation. 
To obtain an averaged azimuthal distribution the bin contents for each shower
are added with the maximum bin aligned  and the azimuthal orientation is
defined according to the direction of  the second highest bin.
The results are shown in figure \ref{fig:phiasymmetry}.
The mean total amplitude of angular  fluctuations in azimuth 
can be as large as $\pm 11\%$ for $10 $\,GeV but decreases approximately with the square root of the primary energy 
to less than $\pm 1.1\,\%$ for $E_0=1 \unit{TeV} $.
However, as indicated by the error bars, the individual bin 
fluctuations are of the same order of magnitude as the mean amplitude of the asymmetry.

The amplitude of the asymmetry is fit with the parabolic function
\begin{equation} \label{eq:azimut:param}
    A\left(\phi\right) = \frac{1}{12}\left[1+b\left\{\left(\phi-\phi_{0}\right)^{2} - \frac{1}{6\pi}\left(\left(2\pi-\phi_{0}\right)^{3}+\phi_{0}^{3}\right) \right\}\right],
\end{equation}
The parameter  $b$ describes the vertical compression and $\phi_0 $ 
the position of the minimum. The third term accounts for the normalization.
The angle $\phi$ is used in units of radians and the bin size of the histogram is $\Delta \phi = 30^\circ = 0.523 \unit{rad}$. 
The results of all fits are given in table 
\ref{tab:fit:asymm} in \ref{app:parares}. 

The parameter $\phi_0$ is found roughly constant. It 
differs slightly from  $\pi $ because the preferential direction 
of the second largest bin leads to an angular bias.
The energy dependency is shown in figure \ref{fig:phiasymmetry} (bottom right). 
The amplitude coefficient $b$  is fit  with 
\eqb \label{eq:energy_dependence_p_asym}
 b =  {p \over \sqrt{E_0}} 
\eqe
as expected from the correspondingly increased number of 
particles in the cascade. 
The results of these fits are summarized in table \ref{tab:fit:p_asymm} in appendix \ref{app:parares}.

\subsection{Dependence of the angular distribution on the shower age \label{sec:agedependence}}

The shape of the angular distribution of tracks $d\hat{l}/d\alpha $ and therefore the angular distribution of \cherl 
has been found to be independent of the energy $E_0$.
However,  the electromagnetic cascade has an  
 extension of a few meter  (see section  \ref{sec:lon}).
Within this evolution of the cascade it is plausible that large 
scattering angles occur more frequently at later stages of the shower development than at 
earlier. Therefore, we investigate how the angular distribution of tracks 
changes with the age of the shower.

\begin{figure}[htp]
    \centering
    \includegraphics*[width=.75\textwidth]{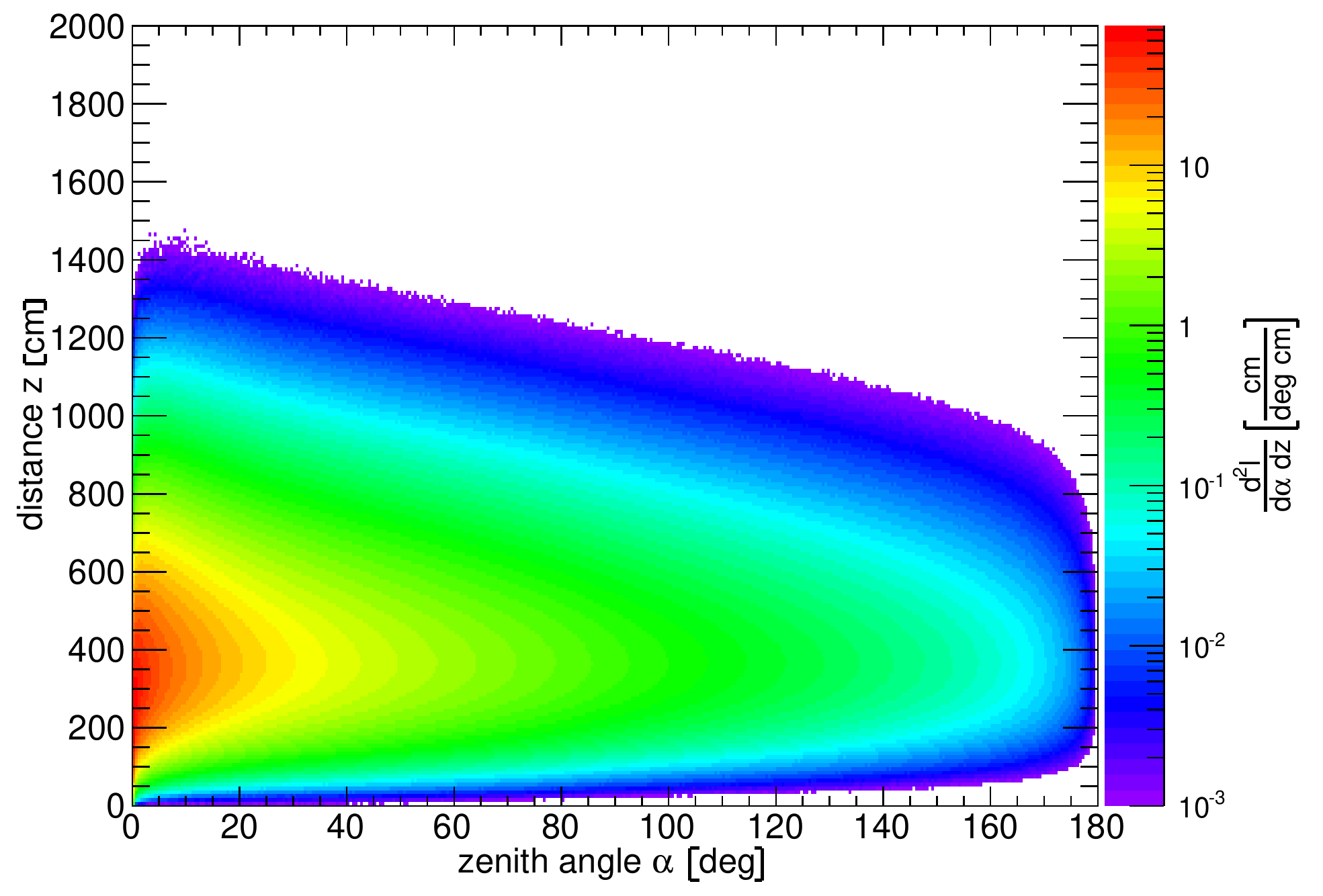}
    \caption{Distribution of the relative track length versus the shower length $z$ and  the inclination angle $\alpha$ for a $1\unit{TeV}$ electron shower. 
The vertical color code corresponds
    to the histogrammed lengths $\hat{l}$ normalized per initial particle.
     \label{fig:zalpha}}
\end{figure}

Figure \ref{fig:zalpha} shows the angular distribution 
of the track length density  versus the longitudinal length of the shower
$ \frac{d^2 \hat{l}}{d \alpha \, d z} $. The distribution
is found to be largely dominated by the longitudinal evolution of the particle density and only a small difference of the angular distribution 
between the onset and the end of the 
cascade can be seen.

\begin{figure}[htp]
    \centering
    \includegraphics*[width=.49\textwidth]{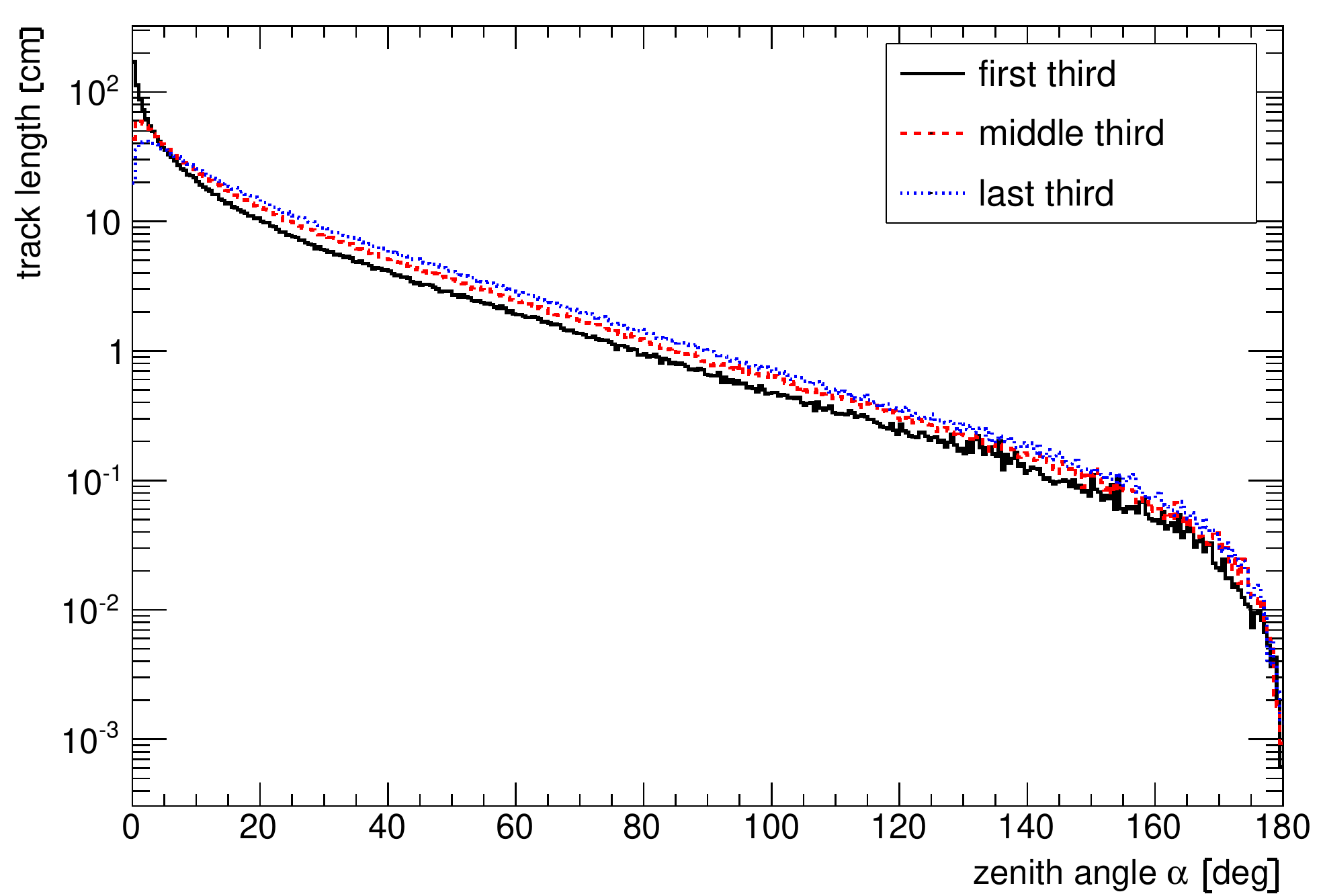}
    \includegraphics*[width=.49\textwidth]{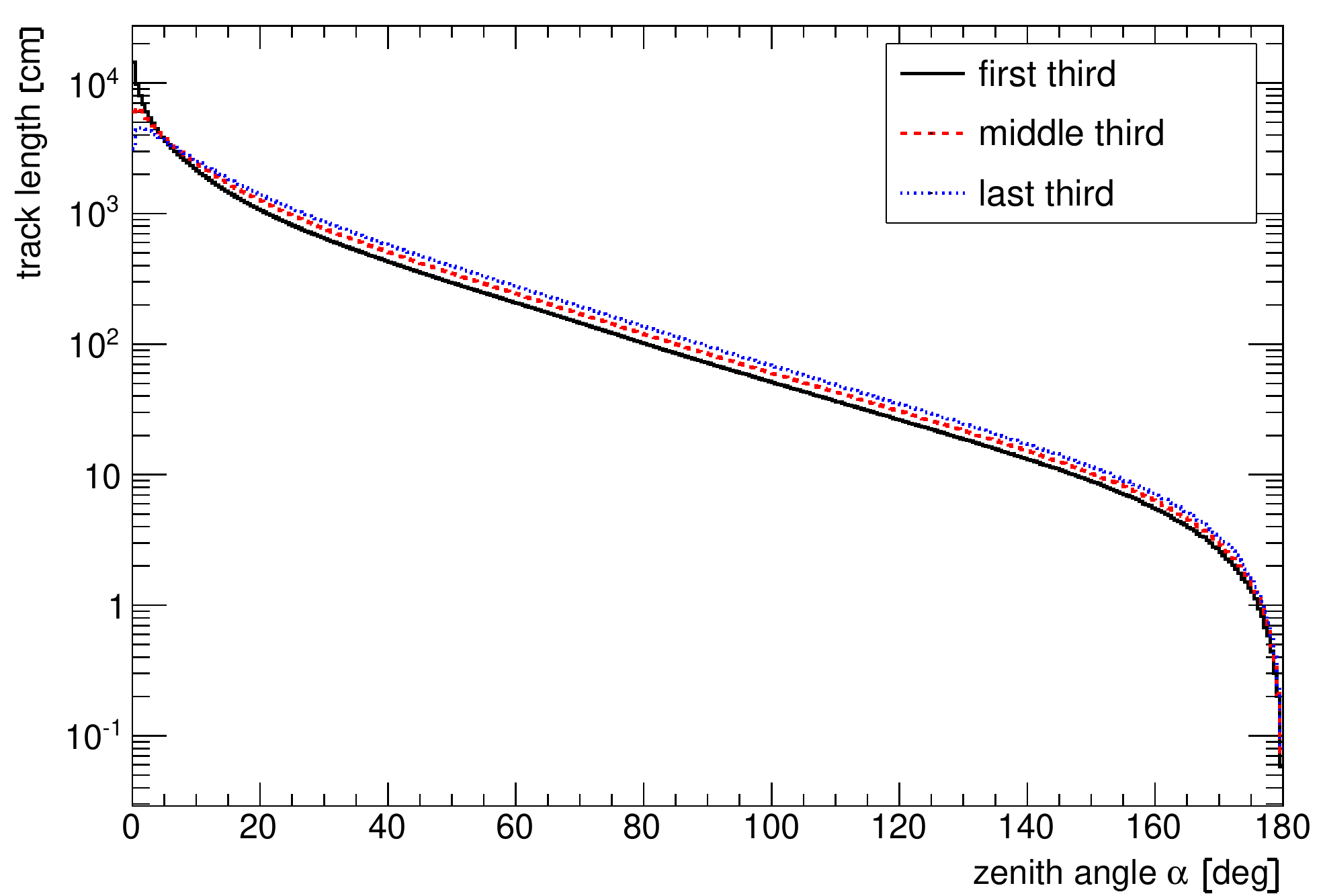}
    \caption{Angular distributions of the track length $\hat{l} $ for different slices of the longitudinal shower evolution.
The figures show the histograms for a primary $\gamma$ with  energy
    $E_0=10\unit{GeV}$ (left) and $E_0=1\unit{TeV}$ (right) normalized
to the track length per particle. The longitudinal distribution has been split into 3 slices of equal total track length.
     \label{fig:alphathirds}}
\end{figure}

For a more detailed  investigation we split cascades 
along the shower axis $z$ into parts of different shower age. We chose 
three slices such that they contain the same total track length and thus emit the same total amount of \cherl.. 
The resulting angular 
distributions are shown in figure \ref{fig:alphathirds}.
Large differences are only seen in the very forward region 
$\alpha < 10^\circ $. 
With increasing
shower age the track length becomes larger  by about a factor $3$.
The differences for large angles $\alpha > 20^\circ $ are comparably smaller,
with about $10\% $ change in yield and no obvious change in shape.
The situation is similar for higher energy $E_0 $.

\begin{figure}[htp]
    \centering
    \includegraphics*[width=.49\textwidth]{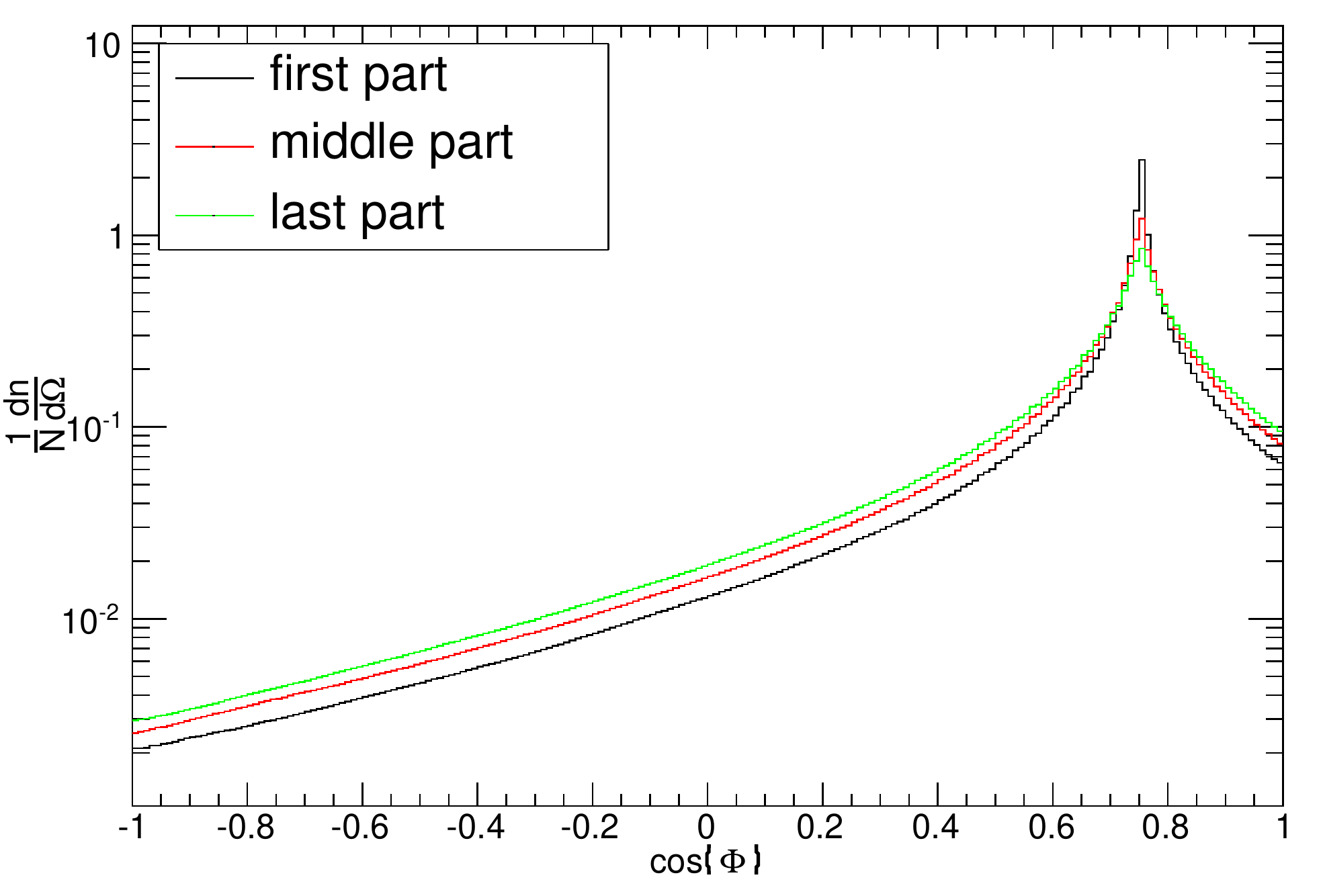}
    \includegraphics*[width=.49\textwidth]{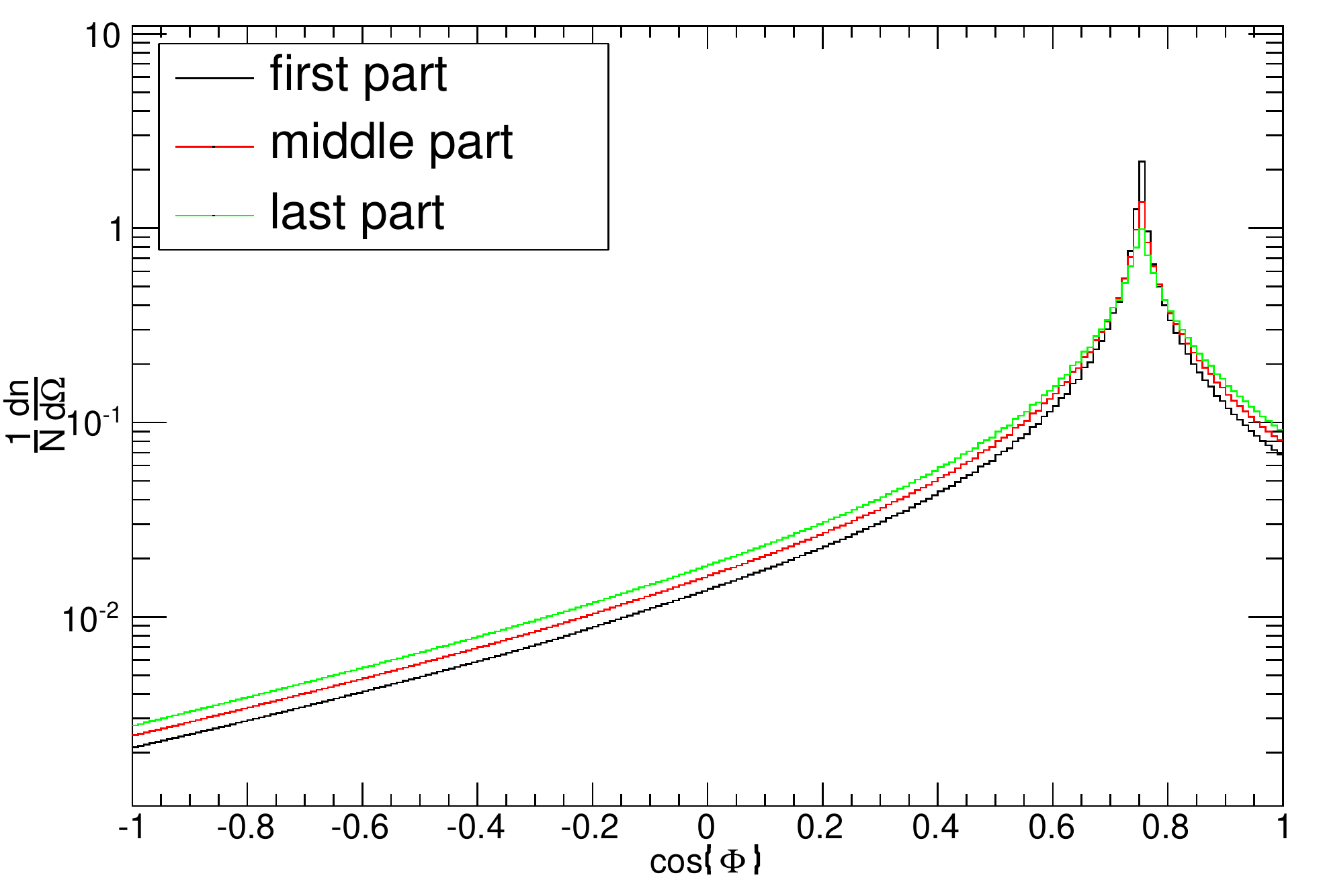}
    \caption{Angular distribution of  the emitted \cherphos 
for the different slices of the longitudinal shower evolution.
The figures show the histograms for a primary $\gamma$ with  energy
    $E_0=10\unit{GeV}$ (left) and $E_0=1\unit{TeV}$ (right) normalized
to the track length per particle. The longitudinal distribution has been split into three slices of equal  track length.
     \label{fig:angulardistthirds}}
\end{figure}

The resulting distributions of \cherphos are shown in figure \ref{fig:angulardistthirds}. The large opening angle of the \cher cone leads to a substantial
 smearing of the angular distribution. Hence, the relatively strong effect into the forward direction for the angular track length distribution does 
not  propagate to an equal strong variation of the \cher peak. Here, only 
an effect of $\pm 10\% - 20 \%$ in the variation of the peak is observed.

In summary, we find that the differential variation of \cherl emission 
along the length of the 
cascade is a relatively weak effect. A global angular distribution, as 
parameterized in section \ref{sec:angular}, seems justified
in particular when considering that the length of a cascade is short 
compared to the typical spacing of optical sensors in neutrino telescopes.
However, for more detailed information we have repeated the 
 angular parameterization Eq.(\ref{eq:angpar:sym}) also 
for the three different slices in shower age separately. The results of the parameterization are given in table \ref{tab:1angcher}, \ref{tab:2angcher} and \ref{tab:2angcher} in \ref{app:parares}.

\section{Summary and Conclusions}

We have simulated electromagnetic cascades with \geantfour for different primary particles and primary energies $E_0$. We  have
parameterized the total Cherenkov-light-radiating track length and its fluctuations, 
the longitudinal
development of the cascade and the angular distribution of 
emitted \cherphos. 

Our result for the total track length agrees within 2\% with the result obtained in \cite{KOWALSKI} but disagrees with other previous calculations.
The relative size of fluctuations for different showers decreases 
$\propto 1/\sqrt{E_0}$. 

The longitudinal profiles are found to be well described by a gamma 
distribution and a difference between $e^\pm$ and 
$\gamma $ is observed as expected \cite{PDG}. However,
quantitatively the position of the shower maximum deviates from the 
values in \cite{PDG}, but agrees with \cite{Grindhammer}. For higher energies than $100 
\unit{GeV} $ we observe a change of slope in the elongation rate.

The  angular distribution of tracks in the cascade and the corresponding distribution of photons is found to be independent of $E_0$  and the type of
primary particle.

Systematic uncertainties of our parameterizations
are  related to the used refraction index and density 
of ice. They are  of the order of $1\%$
for typical values of ice and can be corrected, e.g.  by rescaling 
the observables of our parameterizations which depend on length scales such 
as the axis of the shower development to the correct media density
and by using the correct \cher angle.

Also uncertainties of the used differential 
cross sections of electromagnetic processes 
 in \geantfour can reach up  to  a few percent but are generally substantially smaller
 \cite{GEANT4Physics}. 
We do not expect these differential uncertainties to significantly 
affect our global results, particularly  the total track length is not strongly affected.
Hence we  estimate a typical uncertainty of less than $1\%$.

The LPM effect and dielectric suppression \cite{SPENCERLPM} are considered  
in the used version of \geantfour \cite{schalicke2008improved} but are only expected to become
significant at  larger energies than  considered here.
Electronuclear interactions have not been simulated, because their cross section is small.

The fluctuations of the total track length for different 
individual showers  are found to decrease with
$\propto 1/\sqrt{E_0}$  and the related uncertainty
is already smaller than  $1\% $ for a few GeV.

For the determination of the uncertainties in 
the calculation of the angular distribution of emitted
\cherl we investigate and parameterize the uncertainty related to 
azimuthal fluctuations and the evolution of the angular distribution with increasing shower age.
The azimuthal asymmetry of tracks is found to be small 
(about $7 \% $ at 1 GeV)
and to decrease  $\propto 1/\sqrt{E_0}$. When taking into account  
the large emission angle of \cherphos this effect is expected to
be further washed out for the angular distribution of \cherl 
and is largely negligible at high energies.
Also the effect of the longitudinal 
shower evolution is small and results in differences in  the width 
of the \cher 
peak in the angular distribution. 
This effect becomes  even less important  for  distances 
larger than the  scale length of the cascade 
and it is washed out  e.g.
by the scattering of photons when propagating through  the medium.


\begin{appendix}

\section{\geantfour configuration parameters used for this study \label{app:geant:conf}}

In this chapter a summary of the defined media properties and physics processes is given.

\subsection{Materials}

\begin{table}[htbp]
    \begin{tabular}{lcccc}
        \hline
        \hline
        Medium    & Density $\left[\frac{\mathrm{g}}{\mathrm{cm}^3}\right]$ 
& Index of refraction & Element & Fraction of mass   \\
        \hline
        Ice       & 0.910 & 1.33 & Hydrogen & 88.81\%  \\
                  &       &      & Oxygen   & 11.19\%  \\
        \hline
        \hline
    \end{tabular}
    \caption{Composition of ice as used in the \geantfour-simulation.}
    \label{table:geant:materials}
\end{table}

In \geantfour macroscopic properties of matter are described by \texttt{G4Material} and the atomic
properties  are described by \texttt{G4Element}. A material can consist of multiple
elements and therefore represent a chemical compound, mixture as well as pure materials.
For the performed simulations ice was used.
Unless noted otherwise, the value $n=1.33 $ is used for the 
index of refraction.
The simulated properties of ice are summarized in table \ref{table:geant:materials}.

\subsection{Physicslist}

\begin{table}[htbp]
   \begin{tabular}{llc}
       \hline
       \hline
       Particle  & Process & Model \\
       \hline
       $\gamma$  & G4PhotoElectricEffect & G4PEEffectFluoModel \\
                 & G4ComptonScattering & G4KleinNishinaModel \\
                 & G4GammaConversion  & \\
       \hline
       $e^-$     & G4eMultipleScattering & \\
                 & G4eIonisation & \\
                 & G4eBremsstrahlung & \\
       \hline
       $e^+$     & G4eMultipleScattering & \\
                 & G4eIonisation & \\
                 & G4eBremsstrahlung & \\
                 & G4eplusAnnihilation & \\
       \hline
       $\mu^+,\mu^-$ & G4MuMultipleScattering & \\
                     & G4MuIonisation & \\
                     & G4MuBremsstrahlung & \\
                     & G4MuPairProduction & \\
                     & G4MuNuclearInteraction & \\
                     & G4CoulombScattering & \\
       \hline
       $\pi^+,\pi^-,K^+,K^-,p^+$ & G4hMultipleScattering & \\
                               & G4hIonisation & \\
                               & G4hBremsstrahlung & \\
                               & G4hPairProduction & \\
       \hline
       $\alpha,\mathrm{He}^{3+}$ & G4ionIonisation & \\
                                 & G4hMultipleScattering & \\
                                 & G4NuclearStopping & \\
       \hline
       all unstable particles & G4Decay & \\
       \hline
       \hline
   \end{tabular}
   \caption{Physics processes of most important particles used in the simulation. 
   If no model is specified the default model is used. For hadrons and ions that are
   not listed multiple scattering and ionization are defined.}
   \label{table:geant:physicslist}
\end{table}

All physics processes, which are used during the simulation must be registered in 
\texttt{G4VUserPhysicsList}. These simulations are based on 
the standard physics list \texttt{G4EmStandardPhysics\_option3}.  
The included processes are summarized in table \ref{table:geant:physicslist}.
The  maximum energy for the cross section tables and the 
calculation of  $\mathrm{d}E/\mathrm{d}x$ in
 \geantfour is $10\unit{TeV}$.

\clearpage

\section{Parameterization results \label{app:parares}}

\begin{table}[htbp]
    \centering
    \begin{tabular}{lcc}
        \toprule
\multicolumn{3}{c}{fit of $\hat{l} (E_0 ) $} \\ 
        particle   & $\alpha/\unit{cm\,GeV^{-1}}$     &  $\beta$   \\
        \midrule
        $e^{-}$    & $532.07078881$ & $1.00000211$ \\
        $e^{+}$    & $532.11320598$ & $0.99999254$ \\
        $\gamma$   & $532.08540905$ & $0.99999877$ \\
        \midrule
\multicolumn{3}{c}{fit of $\sigma_{\hat{l}} (E_0 ) $} \\ 
        $e^{-}$    & $5.78170887$ &  \\
        $e^{+}$    & $5.73419669$ & $0.5$ \\
        $\gamma$   & $5.66586567$ &  \\
        \bottomrule
    \end{tabular}
\caption{Result of the parameterization of the effective track length versus primary energy (section \ref{sec:res:total}). \label{tab:result:total}
The top table gives the results of fits of Eq.(\ref{eq:totaltrack}) for the fits
$\hat l (E_0) $ and the bottom the standard deviation $\sigma_{\hat{l}} (E_0 ) $
of the fluctuations.
    \label{tab:results:para_l_sigma_emcascade}}
\end{table}

\begin{table}[htbp]
    \centering
    \begin{tabular}{lccc}
        \toprule
        particle & $\alpha$ & $\beta$ & $b$ \\
        \midrule
        $e^{-}$  & 2.01849 & 1.45469 & 0.63207 \\
        $e^{+}$  & 2.00035 & 1.45501 & 0.63008 \\
        $\gamma$ & 2.83923 & 1.34031 & 0.64526 \\
        \bottomrule
    \end{tabular}
    \caption{Results of the energy dependence of the longitudinal fits, \label{tab:emlong::energypar} section \ref{sec:lon}}
\end{table}

\begin{table}[htbp]
    \centering
    \begin{tabular}{lcFcF}
    \toprule
    particle & $a/\unit{sr^{-1}}$ & \multicolumn{1}{c}{$b$} & $c$ & \multicolumn{1}{c}{$d/\unit{sr^{-1}}$} \\
    \midrule
    $e^{-}$  &  4.27033 & -6.02527 & 0.29887 & -0.00103 \\
    $e^{+}$  &  4.27725 & -6.02430 & 0.29856 & -0.00104 \\
    $\gamma$ &  4.25716 & -6.02421 & 0.29926 & -0.00101 \\
    \bottomrule
    \end{tabular}
    \caption{Averaged  parameters describing the  angular distribution of emitted \cherl, section \ref{sec:angular}.\label{tab:average:angcher}}
\end{table}

\begin{table}[htbp]
\centering
    \begin{longtable}{llcc}
    \toprule
     particle & energy$/\unit{GeV}$ & $a$ & $b$ \\
     \midrule
        $e^{-}$  & 1     & 1.96883 & 0.62794 \\ 
        & 3     & 2.68228 & 0.61705 \\ 
        & 7     & 3.30523 & 0.64303 \\ 
        & 10    & 3.61481 & 0.65247 \\ 
        & 30    & 4.16566 & 0.62248 \\ 
        & 70    & 4.77945 & 0.62823 \\ 
        & 100   & 4.98860 & 0.63416 \\ 
        & 300   & 5.61779 & 0.62033 \\ 
        & 700   & 6.10809 & 0.62129 \\ 
        & 1000  & 6.24439 & 0.61083 \\ 
        & 3000  & 7.03624 & 0.63287 \\ 
        & 7000  & 7.60499 & 0.64319 \\ 
        & 10000 & 7.86789 & 0.65099 \\ 
        \midrule
        $e^{+}$  & 1     & 1.93375 & 0.61535 \\ 
        & 3     & 2.74487 & 0.63234 \\ 
        & 7     & 3.27811 & 0.62541 \\ 
        & 10    & 3.55064 & 0.64072 \\ 
        & 30    & 4.22803 & 0.63283 \\ 
        & 70    & 4.77295 & 0.62917 \\ 
        & 100   & 4.88106 & 0.61512 \\ 
        & 300   & 5.55997 & 0.61605 \\ 
        & 700   & 6.05207 & 0.61644 \\ 
        & 1000  & 6.30800 & 0.62317 \\ 
        & 3000  & 7.01259 & 0.63265 \\ 
        & 7000  & 7.58980 & 0.64096 \\ 
        & 10000 & 7.89891 & 0.65069 \\ 
        \midrule
        $\gamma$  & 1     & 2.49299 & 0.60823 \\ 
        & 3     & 3.66575 & 0.71860 \\ 
        & 7     & 3.99721 & 0.66217 \\ 
        & 10    & 4.10746 & 0.64095 \\ 
        & 30    & 5.08856 & 0.68042 \\ 
        & 70    & 5.33660 & 0.63890 \\ 
        & 100   & 5.60790 & 0.64568 \\ 
        & 300   & 6.08445 & 0.62277 \\ 
        & 700   & 6.61645 & 0.63321 \\ 
        & 1000  & 6.78153 & 0.62575 \\ 
        & 3000  & 7.47467 & 0.64408 \\ 
        & 7000  & 7.96892 & 0.64825 \\ 
        & 10000 & 8.13041 & 0.65240 \\ 
         \bottomrule
    \end{longtable}
    \caption{Results of the  fits \label{tab:emlong:par} of the longitudinal 
    cascade development, section \ref{sec:lon}
    \label{tab:reslon}}
\end{table}

\begin{table}[htbp]
    \centering
    \begin{longtable}{llcFcF}
    \toprule
    particle & energy$/\unit{GeV}$ & $a/\unit{sr^{-1}}$ & \multicolumn{1}{c}{$b$} & $c$ & \multicolumn{1}{c}{$d/\unit{sr^{-1}}$} \\
    \midrule
    $e^{-}$  & 3     &  4.21013 & -6.01199 & 0.30008 & -0.00099 \\
             & 7     &  4.21865 & -6.02752 & 0.30088 & -0.00096 \\
             & 10    &  4.28687 & -6.02147 & 0.29801 & -0.00108 \\
             & 30    &  4.29190 & -6.02932 & 0.29836 & -0.00103 \\
             & 70    &  4.29202 & -6.02913 & 0.29837 & -0.00104 \\
             & 100   &  4.24274 & -6.02275 & 0.29962 & -0.00098 \\
             & 300   &  4.28351 & -6.02684 & 0.29850 & -0.00104 \\
             & 700   &  4.28365 & -6.02672 & 0.29850 & -0.00104 \\
             & 1000  &  4.28351 & -6.02679 & 0.29849 & -0.00103 \\
             & 3000  &  4.28364 & -6.02685 & 0.29850 & -0.00104 \\
             & 7000  &  4.28360 & -6.02690 & 0.29850 & -0.00104 \\
             & 10000 &  4.28365 & -6.02689 & 0.29850 & -0.00103 \\
    \midrule
    $e^{+}$  & 3     &  4.38344 & -6.03618 & 0.29571 & -0.00113 \\
             & 7     &  4.19912 & -6.00949 & 0.30022 & -0.00095 \\
             & 10    &  4.24809 & -6.01802 & 0.29919 & -0.00105 \\
             & 30    &  4.24807 & -6.01820 & 0.29918 & -0.00104 \\
             & 70    &  4.28043 & -6.02557 & 0.29851 & -0.00103 \\
             & 100   &  4.28069 & -6.02577 & 0.29853 & -0.00104 \\
             & 300   &  4.28116 & -6.02627 & 0.29857 & -0.00105 \\
             & 700   &  4.28100 & -6.02623 & 0.29854 & -0.00103 \\
             & 1000  &  4.28106 & -6.02620 & 0.29854 & -0.00103 \\
             & 3000  &  4.28124 & -6.02647 & 0.29856 & -0.00104 \\
             & 7000  &  4.28125 & -6.02656 & 0.29856 & -0.00103 \\
             & 10000 &  4.28134 & -6.02658 & 0.29856 & -0.00103 \\
    \midrule
    $\gamma$ & 3     &  4.15186 & -6.02039 & 0.30274 & -0.00085 \\
             & 7     &  4.14478 & -6.00543 & 0.30202 & -0.00093 \\
             & 10    &  4.30725 & -6.02368 & 0.29745 & -0.00107 \\
             & 30    &  4.31395 & -6.03350 & 0.29786 & -0.00103 \\
             & 70    &  4.25518 & -6.02358 & 0.29922 & -0.00099 \\
             & 100   &  4.25584 & -6.02391 & 0.29928 & -0.00101 \\
             & 300   &  4.25586 & -6.02383 & 0.29926 & -0.00101 \\
             & 700   &  4.25605 & -6.02399 & 0.29926 & -0.00101 \\
             & 1000  &  4.28625 & -6.02812 & 0.29849 & -0.00103 \\
             & 3000  &  4.28632 & -6.02803 & 0.29849 & -0.00103 \\
             & 7000  &  4.28625 & -6.02805 & 0.29848 & -0.00103 \\
             & 10000 &  4.28624 & -6.02800 & 0.29848 & -0.00103 \\
    \bottomrule
    \end{longtable}
\caption{Results of fits of the angular distribution of emitted \cherl, section
\ref{sec:angular}.
\label{tab:angcher}
}
\end{table}

\begin{table}[htbp]
    \centering
    \begin{tabular}{llccc}
    \toprule
    particle & energy$/\unit{GeV}$ & b & $\phi_{0}/\unit{rad}$ & $\sigma $\\
    \midrule
    $e^{-}$  & 1     & 0.07064716 & 3.47699427 & 0.02103594 \\
             & 10    & 0.02131804 & 3.49345733 & 0.00739830 \\
             & 100   & 0.00643688 & 3.52309761 & 0.00234943 \\
             & 1000  & 0.00202350 & 3.48731664 & 0.00074057 \\
             & 10000 & 0.00067586 & 3.46473695 & 0.00023550 \\
    \midrule                                                         
    $e^{+}$  & 1     & 0.07112344 & 3.48404544 & 0.02099490 \\
             & 10    & 0.02097273 & 3.51963176 & 0.00724243 \\
             & 100   & 0.00663420 & 3.48056236 & 0.00232045 \\
             & 1000  & 0.00213402 & 3.46339344 & 0.00073926 \\
             & 10000 & 0.00065039 & 3.50744179 & 0.00023639 \\
    \midrule                                                         
    $\gamma$ & 1     & 0.06652718 & 3.55998809 & 0.02135593 \\
             & 10    & 0.02166958 & 3.47688180 & 0.00730116 \\
             & 100   & 0.00637762 & 3.52556673 & 0.00235480 \\
             & 1000  & 0.00207235 & 3.48507985 & 0.00074559 \\
             & 10000 & 0.00066903 & 3.47379137 & 0.00023858 \\
    \bottomrule
    \end{tabular}
    \caption{Results of fits to the shower asymmetry, Eq.(\ref{eq:azimut:param}) section \ref{sec:azi}.
    The used bin size is $30^\circ $. The column $ \sigma  $
gives the average standard deviation of  fluctuations in each bin.
    \label{tab:fit:asymm}}
\end{table}

\begin{table}[htbp]
    \centering
    \begin{tabular}{lc}
    \toprule
    particle & $p/\unit{rad^{-2}}$ \\
    \midrule
    $e^{-}$  & 0.07029 \\
    $e^{+}$  & 0.07064 \\
    $\gamma$ & 0.06668 \\
    \bottomrule
    \end{tabular}
    \caption{Results of the amplitude coefficient of the shower asymmetry, Eq.(\ref{eq:energy_dependence_p_asym}) section \ref{sec:azi}.
    \label{tab:fit:p_asymm}}
\end{table}

\begin{table}
    \centering
    \begin{longtable}{llcece}
    \toprule
    particle & energy$/\unit{GeV}$ & a & \multicolumn{1}{c}{b} & c & \multicolumn{1}{c}{d} \\
    \midrule
    $e^{-}$  & 3      & 142.9460887 & -9.57395706 & 0.15172749 & -0.00266262 \\
             & 7      & 85.84316237 & -9.05308106 & 0.16313946 & -0.00257812 \\
             & 10     & 77.91266880 & -8.91314499 & 0.16406813 & -0.00284706 \\
             & 30     & 46.82551912 & -8.41395678 & 0.17847633 & -0.00255785 \\
             & 70     & 38.02891569 & -8.19061093 & 0.18434614 & -0.00257831 \\
             & 100    & 34.69834063 & -8.10546380 & 0.18758322 & -0.00247138 \\
             & 300    & 28.97457908 & -7.91471330 & 0.19335737 & -0.00247902 \\
             & 700    & 25.83417400 & -7.79936480 & 0.19747241 & -0.00242657 \\
             & 1000   & 24.40584315 & -7.74365060 & 0.19963483 & -0.00238802 \\
             & 3000   & 23.00296659 & -7.68029984 & 0.20171506 & -0.00238785 \\
             & 7000   & 21.74138458 & -7.62373034 & 0.20390902 & -0.00236181 \\
             & 10000  & 21.46673564 & -7.61084235 & 0.20439010 & -0.00235443 \\
    \midrule
    $e^{+}$  & 3      & 137.9670566 & -9.54748394 & 0.15276498 & -0.00256487 \\
             & 7      & 84.48637367 & -9.00950972 & 0.16249928 & -0.00266916 \\
             & 10     & 71.63054893 & -8.84968017 & 0.16696350 & -0.00265524 \\
             & 30     & 43.90262329 & -8.34648184 & 0.18034774 & -0.00257627 \\
             & 70     & 39.84186034 & -8.23419693 & 0.18270563 & -0.00258497 \\
             & 100    & 34.87615415 & -8.10856986 & 0.18734870 & -0.00249897 \\
             & 300    & 29.07673188 & -7.91861470 & 0.19324693 & -0.00248046 \\
             & 700    & 25.89164868 & -7.80387042 & 0.19747634 & -0.00240558 \\
             & 1000   & 25.19834484 & -7.77379969 & 0.19835006 & -0.00241144 \\
             & 3000   & 23.01752709 & -7.68236819 & 0.20174971 & -0.00237442 \\
             & 7000   & 21.65536829 & -7.62068764 & 0.20410342 & -0.00235278 \\
             & 10000  & 21.33114755 & -7.60423566 & 0.20462732 & -0.00235435 \\
    \midrule
    $\gamma$ & 3      & 69.70469246 & -8.90640987 & 0.17075128 & -0.00211751 \\
             & 7      & 58.54718763 & -8.65787517 & 0.17279750 & -0.00249754 \\
             & 10     & 60.52821188 & -8.64481073 & 0.17011384 & -0.00280772 \\
             & 30     & 46.94065239 & -8.39738538 & 0.17763706 & -0.00264893 \\
             & 70     & 33.61049111 & -8.06982820 & 0.18848567 & -0.00247429 \\
             & 100    & 31.15779106 & -7.99239921 & 0.19102320 & -0.00247773 \\
             & 300    & 26.69846316 & -7.83299913 & 0.19630007 & -0.00244224 \\
             & 700    & 24.57774841 & -7.74984999 & 0.19933205 & -0.00239464 \\
             & 1000   & 23.43128734 & -7.70044416 & 0.20109158 & -0.00238882 \\
             & 3000   & 22.16347238 & -7.64277275 & 0.20314619 & -0.00237112 \\
             & 7000   & 20.95984696 & -7.58781685 & 0.20539395 & -0.00233737 \\
             & 10000  & 20.71724506 & -7.57557901 & 0.20582216 & -0.00233654 \\
    \end{longtable}
    \caption{Results of fits of Eq.(\ref{eq:angpar:sym}) to the angular distribution of \cherl  for the first third of the longitudinal cascade profile, section \ref{sec:agedependence}.
\label{tab:1angcher}}
\end{table}

\begin{table}
    \centering
    \begin{longtable}{llcece}
    \toprule
    particle & energy$/\unit{GeV}$ & a & \multicolumn{1}{c}{b} & c & \multicolumn{1}{c}{d} \\
    \midrule
    $e^{-}$  & 3      & 1.94993953 & -5.57898352 & 0.40890846 & 0.00136720  \\
             & 7      & 2.23159905 & -5.66592786 & 0.38702811 & 0.00096165  \\
             & 10     & 2.41790477 & -5.69781676 & 0.37321829 & 0.00074460  \\
             & 30     & 2.64370991 & -5.72687457 & 0.35830301 & 0.00037598  \\
             & 70     & 2.83551875 & -5.77447099 & 0.34890180 & 0.00017227  \\
             & 100    & 2.87505527 & -5.78793324 & 0.34732224 & 0.00017138  \\
             & 300    & 3.02408620 & -5.81360770 & 0.34020700 & 0.00000261  \\
             & 700    & 3.12021062 & -5.83135934 & 0.33600149 & -0.00008677 \\
             & 1000   & 3.14748514 & -5.83561542 & 0.33478931 & -0.00011422 \\
             & 3000   & 3.22395167 & -5.85249577 & 0.33188936 & -0.00018386 \\
             & 7000   & 3.28080118 & -5.86421471 & 0.32972990 & -0.00022818 \\
             & 10000  & 3.28101811 & -5.86419504 & 0.32973948 & -0.00022808 \\
    \midrule
    $e^{+}$  & 3      & 2.04418647 & -5.57393706 & 0.39850069 & 0.00104086  \\
             & 7      & 2.19550814 & -5.63219580 & 0.38766663 & 0.00100450  \\
             & 10     & 2.38067758 & -5.68162469 & 0.37507854 & 0.00073911  \\
             & 30     & 2.71866151 & -5.75661338 & 0.35520946 & 0.00036425  \\
             & 70     & 2.74306888 & -5.75739258 & 0.35363398 & 0.00031489  \\
             & 100    & 2.85095217 & -5.77142273 & 0.34768022 & 0.00014361  \\
             & 300    & 3.01522146 & -5.81294569 & 0.34066357 & 0.00001562  \\
             & 700    & 3.10293302 & -5.82808660 & 0.33673188 & -0.00006997 \\
             & 1000   & 3.14755994 & -5.83742082 & 0.33491531 & -0.00011240 \\
             & 3000   & 3.21892957 & -5.85126920 & 0.33206727 & -0.00017664 \\
             & 7000   & 3.26963124 & -5.86055633 & 0.33005267 & -0.00022419 \\
             & 10000  & 3.30250950 & -5.86852133 & 0.32892921 & -0.00024216 \\
    \midrule
    $\gamma$ & 3      & 2.31187235 & -5.66208473 & 0.37959321 & 0.00077227  \\
             & 7      & 2.41282189 & -5.68451332 & 0.37267906 & 0.00068690  \\
             & 10     & 2.51707194 & -5.69019396 & 0.36481000 & 0.00052426  \\
             & 30     & 2.75219082 & -5.76825072 & 0.35376221 & 0.00034959  \\
             & 70     & 2.86679434 & -5.77857129 & 0.34716252 & 0.00015492  \\
             & 100    & 2.98152659 & -5.80586065 & 0.34217497 & 0.00003287  \\
             & 300    & 3.06538269 & -5.81769690 & 0.33811151 & -0.00004781 \\
             & 700    & 3.16811393 & -5.84135741 & 0.33410858 & -0.00013899 \\
             & 1000   & 3.19153423 & -5.84578346 & 0.33314079 & -0.00015419 \\
             & 3000   & 3.26243729 & -5.86085963 & 0.33046056 & -0.00020943 \\
             & 7000   & 3.30585081 & -5.86779646 & 0.32870484 & -0.00025242 \\
             & 10000  & 3.30895722 & -5.87088305 & 0.32877273 & -0.00024303 \\
    \bottomrule
    \end{longtable}
    \caption{Results of fits of Eq.(\ref{eq:angpar:sym}) to the angular distribution of \cherl  for the middle third of the longitudinal cascade profile, section \ref{sec:agedependence}.
 \label{tab:2angcher}}
\end{table}

\begin{table}
    \centering
    \begin{longtable}{llcece}
    \toprule
    particle & energy$/\unit{GeV}$ & a & \multicolumn{1}{c}{b} & c & \multicolumn{1}{c}{d} \\
    \midrule
    $e^{-}$  & 3      & 142.6044855 & -9.57205854 & 0.15179690 & -0.00265801 \\
             & 7      & 85.99022343 & -9.05411822 & 0.16307753 & -0.00257921 \\
             & 10     & 77.79962657 & -8.91179784 & 0.16410414 & -0.00284619 \\
             & 30     & 46.88367741 & -8.41476975 & 0.17842752 & -0.00255845 \\
             & 70     & 38.08621359 & -8.19192075 & 0.18429858 & -0.00257872 \\
             & 100    & 34.72957609 & -8.10617168 & 0.18754369 & -0.00247331 \\
             & 300    & 28.97315954 & -7.91468463 & 0.19335970 & -0.00247882 \\
             & 700    & 25.95546681 & -7.80308554 & 0.19724201 & -0.00243642 \\
             & 1000   & 24.48558230 & -7.74579464 & 0.19946929 & -0.00239809 \\
             & 3000   & 23.04520886 & -7.68159582 & 0.20162205 & -0.00239248 \\
             & 7000   & 21.71336645 & -7.62283497 & 0.20397007 & -0.00235902 \\
             & 10000  & 21.46127003 & -7.61065501 & 0.20440436 & -0.00235375 \\
    \midrule
    $e^{+}$  & 3      & 138.1008936 & -9.54847446 & 0.15274270 & -0.00256489 \\
             & 7      & 84.49824682 & -9.00981190 & 0.16249033 & -0.00266739 \\
             & 10     & 71.69254483 & -8.85031418 & 0.16693985 & -0.00265661 \\
             & 30     & 43.95573617 & -8.34754329 & 0.18030217 & -0.00257905 \\
             & 70     & 39.85957512 & -8.23454455 & 0.18268712 & -0.00258586 \\
             & 100    & 34.87521933 & -8.10854937 & 0.18735001 & -0.00249890 \\
             & 300    & 29.07085848 & -7.91850474 & 0.19325786 & -0.00247968 \\
             & 700    & 25.88875951 & -7.80377086 & 0.19748135 & -0.00240549 \\
             & 1000   & 25.19809801 & -7.77379079 & 0.19835066 & -0.00241142 \\
             & 3000   & 23.01750888 & -7.68236750 & 0.20174975 & -0.00237442 \\
             & 7000   & 21.65536329 & -7.62068747 & 0.20410343 & -0.00235278 \\
             & 10000  & 21.33114596 & -7.60423560 & 0.20462733 & -0.00235435 \\
    \midrule
    $\gamma$ & 3      & 69.85074756 & -8.90835590 & 0.17069329 & -0.00211976 \\
             & 7      & 58.28180667 & -8.65424155 & 0.17295947 & -0.00248847 \\
             & 10     & 60.49942091 & -8.64435137 & 0.17012934 & -0.00280746 \\
             & 30     & 47.11516858 & -8.40045586 & 0.17750367 & -0.00265504 \\
             & 70     & 33.55526488 & -8.06827181 & 0.18855456 & -0.00247353 \\
             & 100    & 31.16182031 & -7.99249181 & 0.19101555 & -0.00247788 \\
             & 300    & 26.71881146 & -7.83341114 & 0.19625992 & -0.00244509 \\
             & 700    & 24.56636625 & -7.74960945 & 0.19935804 & -0.00239229 \\
             & 1000   & 23.43436187 & -7.70054100 & 0.20108516 & -0.00238910 \\
             & 3000   & 22.16109501 & -7.64269786 & 0.20315149 & -0.00237087 \\
             & 7000   & 20.96330326 & -7.58792493 & 0.20538521 & -0.00233778 \\
             & 10000  & 20.71783561 & -7.57559703 & 0.20582042 & -0.00233665 \\
    \bottomrule
    \end{longtable}
    \caption{Results of fits of Eq.(\ref{eq:angpar:sym}) to the angular distribution of \cherl  for the last third of the longitudinal cascade profile, section \ref{sec:agedependence}.
\label{tab:3angcher}}
\end{table}

\clearpage

\end{appendix}

\section*{Acknowledgement}

We  thank the IceCube group at the RWTH Aachen University for fruitful 
discussions.
We thank Dmitry Chirkin and Spencer Klein for reading the manuscript and 
valuable suggestions.
This work is supported by the German Ministry for Education and Research (BMBF).

\bibliographystyle{elsarticle-num}
\bibliography{references}

\end{document}